\documentclass[twocolumn]{aastex62}
\usepackage{color,soul,appendix}

\PassOptionsToPackage{draft}{hyperref}
\usepackage{xspace}

\usepackage[xindy, toc, hyperfirst=false, nolist, nostyles, sanitize={name=false,description=false,symbol=false}]{glossaries}
\glsdisablehyper
\usepackage[hyperref,x11names, table]{}

\newglossaryentry{vrad}{name={radial velocity~}, text={radial velocity}, symbol={\ensuremath{v_\textrm{rad}}}, description={radial velocity}, sort=vrad}
\newglossaryentry{vrot}{name={stellar rotation~}, name={stellar rotation}, symbol={\ensuremath{v_\textrm{rot}}}, description={radial velocity}, sort=vrot}

\newcommand{\xray}{X-ray}

\newglossaryentry{angstrom}{name=\AA, description={unit of length $10^{-10}$\,m}, sort=angstrom}
\newglossaryentry{nir}{name=NIR,description={near infrared},first = {near infrared (NIR)}}
\newglossaryentry{psf}{name=PSF,description={point-spread function},first = {point-spread function (PSF)}}
\newglossaryentry{fwhm}{name=FWHM,description={Full Width Half Maximum},first = {FWHM}}
\newglossaryentry{rms}{name=RMS,description={Root Mean Square},first = {RMS}}
\newglossaryentry{signalnoise}{name=S/N,description={signal to noise}}
\newglossaryentry{uv}{name=UV,description={ultra violet},first = {ultra violet (UV)}}
\newglossaryentry{halpha}{name=\ensuremath{\textrm{H}\alpha}, description={First line of the Balmer series at 6563\,\AA}, sort=halpha}
\newglossaryentry{mgb}{name={Mg \textsc{i} b}, description={Triplet at 5167\,\AA, 5173\,\AA and 5184\,\AA}}
\newglossaryentry{sobolevapprox}{name={Sobolev approximation}, description={Lines are approximation with an infinitley thin interaction region \citep[e.g. no broadening][]{1960mes..book.....S}}, first={Sobolev approximation }}
\newglossaryentry{radeq}{name={radiative equilibrium}, description={The net flux of energy between matter and radiation field is zero}}
\newglossaryentry{nebularapprox}{name={nebular approximation}, description={Assumes that the plasma condition are controlled by a central radiation source. The radiation field decreases with the distance to the source by geometrical dilution. See \citet{1978stat.book.....M} for details}}
\newglossaryentry{modnebularapprox}{name={modified nebular approximation}, description={In contrast to \gls{nebularapprox} where only geometrical dilution is taken into account, the modified nebular approximation also takes dilution by other radiative processes into account }, first={modified nebular approximation}, parent=nebularapprox}
\newglossaryentry{thompsonscat}{name={Thomson scattering}, description={Scattering of photons on low energy electrons}}
\newglossaryentry{lte}{name={LTE}, description={Local Thermodynamic Equilibrium}, first={local thermodynamic equilibrium (LTE)}}
\newglossaryentry{lsr}{name={LSR}, description={Local Standard of Rest}, first={\textit{local standard of rest} (LSR)}}
\newglossaryentry{mc}{name={MC}, description={Monte Carlo}, first={\textit{Monte Carlo} (MC)}}
\newglossaryentry{wcs}{name={WCS}, description={world coordinate system}, first={world coordinate system (WCS)}}
\newglossaryentry{cmf}{name=CMF, text=CMF, first=Comoving Frame (CMF henceforth), description={Comoving Frame}}

\newglossaryentry{uvoir}{name=UVOIR, text=UVOIR, first=UV/optical/Near-IR (UVOIR), description={UV/optical/Near-IR}}

\newglossaryentry{sfit}{name=SFIT, text=\textsc{sfit}, description={spectral fitting program for hot stars \citep{2001A&A...376..497J}}, first={\textsc{sfit} \citep{2001A&A...376..497J}}}
\newglossaryentry{iraf}{name=IRAF, text=\textsc{iraf}, description={Image Reduction and Analysis Facility maintained by NOAO}, first={\textsc{iraf}\protect\footnote{IRAF: the Image Reduction and Analysis Facility is distributed by the National Optical Astronomy Observatory, which is operated by the Association of Universities for Research in Astronomy (AURA) under cooperative agreement with the National Science Foundation (NSF).}}}
\newglossaryentry{pyraf}{name=PyRAF, text=\textsc{PyRAF}, description={Python wrap of \gls{iraf} maintained by STSCI}, first=\textsc{PyRAF} \protect\footnote{PyRAF is a product of the Space Telescope Science Institute, which is operated by AURA for NASA.}}
\newglossaryentry{astropy}{name=ASTROPY, text=\textsc{astropy}, description=\textsc{astropy} framework, first = \textsc{astropy} \citep{2013A&A...558A..33A}}
\newglossaryentry{numpy}{name=NUMPY, text=\textsc{numpy}, description=\textsc{numpy} framework, first = \textsc{numpy} \citep{walt2011numpy}}
\newglossaryentry{scipy}{name=SCIPY, text=\textsc{scipy}, description=\textsc{scipy} framework, first = \textsc{scipy} \citep{Jones:2001fk}}
\newglossaryentry{matplotlib}{name=matplotlib, text=\textsc{matplotlib}, description=\textsc{matplotlib} framework, first = \textsc{matplotlib} \citep{hunter2007matplotlib}}
\newglossaryentry{pandas}{name=pandas, text=\textsc{pandas}, description=\textsc{pandas} framework, first = \textsc{pandas} \citep{mckinney2010data}}
\newglossaryentry{ipython}{name=ipython, text=\textsc{ipython}, description=\textsc{ipython} framework, first = \textsc{ipython} \citep{perez2007ipython}}
\newglossaryentry{jupyter}{name=jupyter, text=\textsc{jupyter}, description=\textsc{jupyter} framework, first = \textsc{jupyter} \citep{kluyver2016jupyter,perez2015project,ragan2014jupyter}}
\newglossaryentry{aplpy}{name=aplpy, text=\textsc{aplpy}, description=\textsc{aplpy} framework, first = \textsc{aplpy} \citep{2012ascl.soft08017R}}
\newglossaryentry{nltk}{name=nltk, text=\textsc{nltk}, description=\textsc{nltk} framework, first = Natural Language ToolKit \citep[\textsc{NLTK};][]{bird2009natural}}
\newglossaryentry{scikit-learn}{name=scikit-learn, text=\textsc{scikit-learn}, description=\textsc{scikit-learn} framework, first = \textsc{scikit-learn} \citep[][]{scikit-learn}}
\newglossaryentry{scikit-image}{name=scikit-image, text=\textsc{scikit-image}, description=\textsc{scikit-image} framework, first = \textsc{scikit-image} \citep[][]{scikit-image}}
\newglossaryentry{moog}{name=MOOG,text={\textsc{moog}}, description={spectral synthesis software \citep{1973ApJ...184..839S}}, first={\textsc{Moog} \citep{1973ApJ...184..839S}}}
\newglossaryentry{atlas9}{name=ATLAS9,description={grid of stellar atmospheres \citep{2004astro.ph..5087C}}, first={ATLAS9 \citep{2004astro.ph..5087C}}}
\newglossaryentry{vald}{name=VALD,description={Vienna Atomic Line Database \citep{2000BaltA...9..590K}}, first={Vienna Atomic Line Database \citep[VALD;][]{2000BaltA...9..590K}}}
\newglossaryentry{sextractor}{name=SExtractor, text=\textsc{SExtractor}, description={Source Extractor photometry program \citep{1996A&AS..117..393B}}, first={\textsc{SExtractor} \citep{1996A&AS..117..393B}}}

\newglossaryentry{swarp}{name=SWarp, text=\textsc{SWarp}, description={SWarp \citep{2002ASPC..281..228B}}, first={\textsc{SWarp} \citep{2002ASPC..281..228B}}}
\newglossaryentry{astrometry.net}{name=astrometry.net, text=\textsc{astrometry.net}, description={\textsc{astrometry.net} \citep{2010AJ....139.1782L}} first={\textsc{astrometry.net} \citep{2010AJ....139.1782L}}}

\newglossaryentry{astrodrizzle}{name=AstroDrizzle, text=\textsc{AstroDrizzle}, description={AstroDrizzle \citep{2012drzp.book.....G}}, first={\textsc{AstroDrizzle} \citep{2012drzp.book.....G}}}

\newglossaryentry{idl}{name=IDL,text={\textsc{idl}}, description={Interactive Data Language}}
\newglossaryentry{makee}{name=MAKEE,text=\textsc{makee}, description={MAuna Kea Echelle Extraction by Tom Barlow available}}% at \verb+http://spider.ipac.caltech.edu/staff/tab/makee/index.html+}}
\newglossaryentry{minuit}{name=MINUIT,text={\textsc{minuit}}, description={collection of numerical optimization tools \citep{James:1975dr}}}
\newglossaryentry{migrad}{name=MIGRAD,text={\textsc{migrad}}, description={numerical gradient optimization tools - part of \gls{minuit}}}
\newglossaryentry{dolphot}{name=DOLPHOT, text=\textsc{dolphot}, description=photometry package for HST, first=\textsc{dolphot} \citep{2000PASP..112.1383D}}
\newglossaryentry{synphot}{name=synphot, text={\textsc{synphot}}, description={synthetic photometry package from STSCI}, first={\textsc{synphot}\protect\footnote{\textsc{synphot} is a product of the Space Telescope Science Institute, which is operated by AURA for NASA.}}}
\newglossaryentry{chianti}{name=CHIANTI, text=CHIANTI, description= CHIANTI Database 7.1, first =CHIANTI 7.1 \citep{1997A&AS..125..149D,2012ApJ...744...99L}}
\newglossaryentry{synpp}{name=SYNPP, text=SYN++, description= SYN++ software, first =SYN++ \citep{2011PASP..123..237T}}
\newglossaryentry{tardis}{name=TARDIS, text=\textsc{tardis}, description= TARDIS MC code, first = {\textsc{tardis} \citep{2014MNRAS.440..387K,kerzendorf_wolfgang_2022_6662839}}}
\newglossaryentry{hach12}{name=H12, text=H12, description= Hachinger et al 2012 paper on 94I and hidden He, first = {\citep[][henceforth referred to as H12]{hachinger94Imodel}}}

\newglossaryentry{snid}{name=SNID, text=\textsc{snid}, description= Supernova Identification code, first={\textsc{snid} \citep{snid}}}
\newglossaryentry{lcogt}{name=LCO, text=LCO, description= LCO, first = {LCO \citep{lcogt}}}
\newglossaryentry{lasair}{name=LASAIR, text=LASAIR, description= LASAIR, first = {LASAIR \citep{lasair}}}

\newglossaryentry{artis}{name=ARTIS, text=\textsc{artis}, description= ARTIS MC code, first = \textsc{artis} \citep{2009MNRAS.398.1809K}}
\newglossaryentry{cmfgen}{name=CMFGEN, text=\textsc{cmfgen}, description=CMFGGEn radiative transfer code, first = \textsc{cmfgen} \citep{1998ApJ...496..407H}}
\newglossaryentry{sedona}{name=SEDONA, text=\textsc{sedona}, description= Sedona MC code, first = \textsc{sedona} \citep{2006ApJ...651..366K}}
\newglossaryentry{phoenix}{name=PHOENIX, text=\textsc{phoenix}, description= PHOENIX radiative transfer code, first = \textsc{phoenix} \citep{1997ApJ...490..803H}}
\newglossaryentry{mlmc}{name=MLMC, text=ML93, description= Mazzali Lucy Monte Carlo, first ={Mazzali \& Lucy (1993, ML93) code}}
\newglossaryentry{starkit}{name=STARKIT, text=\textsc{starkit}, description= TARDIS MC code, first = {\textsc{starkit} \citep{wolfgang_kerzendorf_2015_28016}}}

\newglossaryentry{pyne}{name=PYNE, text=\textsc{pyne}, description= PYNE code, first = {\textsc{pyne} \citep{Scopatz2012a}}}
\newglossaryentry{multinest}{name=MULTINEST, text=\textsc{MultiNest}, description=MultiNest, first={\textsc{MultiNest} \citep{2009MNRAS.398.1601F}}}
\newglossaryentry{wsynphot}{name=WSYNPHOT, text=\textsc{wsynphot}, description=Wsynphot, first={\textsc{wsynphot}\protect\footnote{\protect\url{https://github.com/wkerzendorf/wsynphot}}}}
\newglossaryentry{specutils}{name=SPECUTILS, text=\textsc{specutils}, description=specutils, first={\textsc{specutils} \protect\footnote{\protect\url{https://github.com/astropy/specutils}}}}
\newglossaryentry{ads}{name=ADS ,description=ADS, first={NASA Astrophysics Data System (ADS) \citep{2000A&AS..143...41K}}}

\newglossaryentry{2mass}{name=2MASS,description={Two Micron All Sky Survey \citep{2006AJ....131.1163S}}, first={Two Micron All Sky Survey \citep{2006AJ....131.1163S}}}
\newglossaryentry{wiserep}{name=\textsc{WISeREP}, description={Weizmann Interactive Supernova data REPository \citep{2006AJ....131.1163S}}, first={\textsc{WISeREP} \citep{2012PASP..124..668Y}}}
\newglossaryentry{nomad}{name=NOMAD,first={Naval Observatory Merged Astrometric Dataset \citep[NOMAD; ][]{2005yCat.1297....0Z}}, description={Naval Observatory Merged Astrometric Dataset}}
\newglossaryentry{wifes}{name=WIFES, text=\textsc{WiFeS}, first={\textsc{WiFeS} \citep{2007Ap&SS.310..255D}},  description={Wide Field Spectrograph - \gls{ifu} mounted on the 2.3\,m telescope at Siding Spring Observatory}}
\newglossaryentry{scp}{name=SCP,description={Supernova Cosmology Project, led by Saul Perlmutter}, first={Supernova Cosmology Project (SCP)}}
\newglossaryentry{hzsns}{name=HZSNS,description={High Z Supernova Search, led by Brian Schmidt}, first={High Z Supernova Search (HZSNS)}}
\newglossaryentry{vlt}{name=VLT,description={Very Large Telescope located on Cerro Paranal (Chile)}, first={Very Large Telescope (VLT)}}
\newglossaryentry{flames}{name=FLAMES,description={Multi-object, intermediate and high resolution spectrograph mounted on the  \gls{vlt}}}
\newglossaryentry{hires}{name=HIRES, description={High Resolution Echelle Spectrometer mounted on the Keck Telescope}, first={High Resolution Echelle Spectrometer \citep[HIRES;][]{1994SPIE.2198..362V}}}
\newglossaryentry{lris}{name=LRIS,description={Low Resolution Imaging Spectrometer mounted on the Keck Telescope}, first={Low-Resolution Imaging Spectrometer \citep[LRIS;][]{Oke95}}}

\newglossaryentry{decam}{name=DECam, description={DECam is a high-performance, wide-field CCD imager mounted at the prime focus of the Blanco 4-m telescope at \gls{ctio}.}, first={Dark Energy Camera \citep[DECam; ][]{2012PhPro..37.1332D,2015AJ....150..150F}}}

\newglossaryentry{essence}{name=ESSENCE,description={The `Equation of State: SupErNovae trace Cosmic Expansion' project \citep[ESSENCE;][]{2002AAS...201.7809G}}, first={`The Equation of State: SupErNovae trace Cosmic Expansion' \citep[ESSENCE;][]{2002AAS...201.7809G}}}
\newglossaryentry{ifu}{name=IFU,description={Optical instrument combining spectrographic and imaging capabilities, used to obtain spatially resolved spectra}, first={Integral Field Unit (IFU)}, firstplural={Integral Field Units (IFUs)}}

\newglossaryentry{besancon}{name=Besan\c{c}on Model, description={Model of stellar population synthesis of the Galaxy, including kinematics.}}%  \verb+http://model.obs-besancon.fr+} }, nonumberlist=true}

\newglossaryentry{int}{name=INT,description={Isaac Newton 2.5\,m Telescope}, first={Isaac Newton 2.5\,m Telescope (INT)}}
\newglossaryentry{iau}{name=IAU,description={International Astronomical Union}, first={IAU}}
\newglossaryentry{chandra}{name=Chandra,description={Chandra \xray\ Observatory (space-based)}}
\newglossaryentry{hst}{name=HST,description={Hubble Space Telescope}}
\newglossaryentry{hst.wfpc2}{name=WFPC2,description={Wide-Field Planetary Camera 2 mounted on the \gls{hst}}, first={Wide-Field Planetary Camera 2 (WFPC2)}}
\newglossaryentry{hst.acs}{name=ACS,description={Advanced Camera for Surveys mounted on the \gls{hst}}, first={Advanced Camera for Surveys (ACS)}}
\newglossaryentry{hst.wfc3}{name=WFC3,description={Wide-Field Camera 3 mounted on the \gls{hst}}, first={Wide-Field Camera 3 (WFC3)}}
\newglossaryentry{hst.cte}{name=CTE, description={charge transfer efficiency (CTE)}, first={charge transfer efficiency \citep[CTE; see ][for a description]{2009acs..rept....1C}}}

\newglossaryentry{snls}{name=SNLS,description={Supernova Legacy Survey \citep{2003AAS...203.8209P}}, first={Supernova Legacy Survey \citep[SNLS;][]{2003AAS...203.8209P}}}
\newglossaryentry{dass}{name=DASS, description={Digitized Astronomy Supernova Survey \citep{1975PASP...87..565C}}, first={Digitized Astronomy Supernova Survey \citep[DASS;][]{1975PASP...87..565C}}}
\newglossaryentry{bait}{name=BAIT, description={Berkley Automatic Imaging Telescope \citep{1993PASP..105.1164R}}, first={Berkley Automatic Imaging Telescope \citep[BAIT;][]{1993PASP..105.1164R}}}
\newglossaryentry{kait}{name=KAIT, description={Katzman Automatic Imaging Telescope \citep{2001ASPC..246..121F}}, first={Katzman Automatic Imaging Telescope \citep[KAIT;][]{2001ASPC..246..121F}}}
\newglossaryentry{loss}{name=LOSS, description={Lick Observatory Supernova Search  \citep{2000AIPC..522..103L}}, first={Lick Observatory Supernova Search \citep[LOSS;][]{2000AIPC..522..103L}}}
\newglossaryentry{ctss}{name=CTSS,description={Cal\'{a}n/Tololo Supernova Survey \citep{1993AJ....106.2392H}}, first={Cal\'{a}n/Tololo supernova survey \citep[CTSS;][]{1993AJ....106.2392H}}}
\newglossaryentry{ctio}{name= CTIO, description={Cerro Tololo Inter-American Observatory}, first={Cerro Tololo Inter-American Observatory (CTIO)}}
\newglossaryentry{ptf}{name=PTF, description={Palomar Transient Factory \citep{2009PASP..121.1334R}}, first={Palomar Transient Factory \citep[PTF;][]{2009PASP..121.1334R}}}
\newglossaryentry{batse}{name=BATSE, description={Burst and Transient Source Experiment mounted on the Compton Gamma Ray Observatory}, first={Burst and Transient Source Experiment (BATSE)}}
\newglossaryentry{bepposax}{name=BeppoSAX, description={\xray\ satellite named in honor of Giuseppe "Beppo" Occhialini}}
\newglossaryentry{rosat}{name=ROSAT, description={short for R\"{o}ntgensatellit}, first={ROSAT}}
\newglossaryentry{hete2}{name=HETE2, description={High Energy Transient Explorer}, first={High Energy Transient Explorer (HETE)}}
\newglossaryentry{ska}{name=SKA, description={Square Kilometre Array}, first={Square Kilometre Array (SKA)}}

\newglossaryentry{gnirs}{name=GNIRS, description={Gemini Near InfraRed Spectrograph mounted on the Gemini North Telescope}}
\newglossaryentry{gmosn}{name=GMOS, description={Gemini Multi Object Spectrograph mounted on the
 Gemini North Telescope}, first={GMOS \citep[Gemini Multi Object Spectrograph;][]{2004PASP..116..425H}}}
\newglossaryentry{swift}{name=Swift, description={Swift Gamma-Ray Burst Mission}}
\newglossaryentry{vla}{name=VLA, description={Very Large Array radio telescope located in North America}, first={Very Large Array (VLA)}}
\newglossaryentry{evla}{name=EVLA, description={Extended Very Large Array radio telescope located in North America}, first={Extended Very Large Array (EVLA)}}
\newglossaryentry{sdss}{name=SDSS, description={Sloan Digital Sky Survey}}
\newglossaryentry{dss}{name=DSS, description={Digitized Sky Survey}}
\newglossaryentry{skymapper}{name=SkyMapper, description={SkyMapper telescope \citep{2007PASA...24....1K}}, first={SkyMapper \citep{2007PASA...24....1K}}}
\newglossaryentry{panstarrs}{name=PanSTARRS, description={Panoramic Survey Telescope \& Rapid Response System \citep{2004SPIE.5489...11K}}, first={Panoramic Survey Telescope \& Rapid Response System \citep[PanSTARRS;][]{2004SPIE.5489...11K}}}
\newglossaryentry{ps1dr1}{name=PS1~DR1, description={Panoramic Survey Telescope \& Rapid Response System \citep{2004SPIE.5489...11K} }, first={Panoramic Survey Telescope \& Rapid Response System \citep[PanSTARRS;][]{2004SPIE.5489...11K} DR1}}

\newglossaryentry{lsst}{name=LSST, description={Large Synoptic Survey Telescope}, first={Large Synoptic Survey Telescope \citep[LSST;][]{2006AAS...209.8604P}}}
\newglossaryentry{ppmxl}{name=PPMXL, description={PPMXL Catalog of Positions and Proper Motions on the ICRS \citep{2010AJ....139.2440R}}}
\newglossaryentry{gaia}{name=GAIA, description={Global Astrometric Interferometer for Astrophysics \citep{2001A&A...369..339P}}, first={Global Astrometric Interferometer for Astrophysics \citep[GAIA;][]{2001A&A...369..339P}}}
\newglossaryentry{ligo}{name=LIGO, description={Laser Interferometer Gravitational Wave Observatory}, first={Laser Interferometer Gravitational Wave Observatory \citep[LIGO;][]{1992Sci...256..325A}}}
\newglossaryentry{aligo}{name=Advanced LIGO, description={Advanced LIGO}, sort=ligo2}
\newglossaryentry{lisa}{name=LISA, description={Laser Interferometer Space Antenna \citep{1994ESAJ...18..219J}}, first={Laser Interferometer Space Antenna \citep[LISA;][]{1994ESAJ...18..219J}}}
\newglossaryentry{ccd}{name=CCD,description={Charged Coupled Device}, first={charged coupled device (CCD)}, firstplural={charged coupled devices (CCDs)}}

\newcommand{\sn}[2]{SN~#1#2\xspace}

\newglossaryentry{irc}{name=IRC, text={IRC}, description={infrared catastrophe}, first={infrared catastrophe \citep[IRC;][]{1980PhDT.........1A}}}

\newglossaryentry{sn}{name=Supernova, text={SN}, plural={SNe}, description={exploding star}, nonumberlist=true, first={supernova (SN)}, firstplural={supernovae (SNe)}}
\newglossaryentry{snia}{name=Type~Ia (SN~Ia), text={SN~Ia}, description={Thermonuclear explosion of a white dwarf - spectra show no hydrogen but a strong silicon line},first={Type~Ia supernova (SN~Ia)}, firstplural={Type Ia supernovae (SNe~Ia)}, plural={SNe~Ia}, parent=sn, nonumberlist=true}
\newcommand{\sneia}{\glspl*{snia}\xspace}

\newglossaryentry{branchnormal}{name={branch-normal}, text=\textit{Branch-normal}, description={Large homogeneous class of Type Ia Supernovae, defined in \citet{1993AJ....106.2383B}}, first={\textit{Branch-normal} SNe Ia \citep{1993AJ....106.2383B}}, parent=snia}
\newglossaryentry{91t}{name={91T-like}, description={Luminous class of Type Ia supernovae similar to \sn{1991}{T} \citep{1992AJ....103.1632P}} , first={91T-like}, parent=snia}
\newglossaryentry{91bg}{name={91bg-like}, description={Faint class of Type Ia supernovae similar to \sn{1991}{bg} \citep{1992AJ....104.1543F}}, first={91bg-like}, parent=snia}
\newglossaryentry{02cx}{name={02cx-like}, description={Peculiar class of Type Ia supernovae similar to \sn{2002}{cx} \citep{2003PASP..115..453L}}, first={02cx-like \sneia\ \citep{2003PASP..115..453L}}, parent=snia}

\newglossaryentry{snibc}{name=Type~Ib/c, text={SN~Ib/c}, description={Collapse of the core of a massive star -  spectrum shows no hydrogen and no silicon line},first={Type~Ib/c supernova (SN~Ib/c)}, firstplural={Type~Ib/c supernovae (SNe~Ib/c)}, plural={SNe~Ib/c}, parent=sn}

\newglossaryentry{snib}{name=Type~Ib, text={SN~Ib}, description={Spectrum shows no hydrogen and no silicon, but helium line},first={Type Ib supernova (SN~Ib)}, firstplural={Type~Ib supernovae (SNe~Ib)}, plural={SNe~Ib}, parent=snibc}

\newglossaryentry{snic}{name=Type~Ic, text={SN~Ic}, description={Spectrum shows no hydrogen, no silicon and no helium line},first={Type~Ic supernova (SN~Ic)}, firstplural={Type~Ic supernovae (SNe~Ic)}, plural={SNe~Ic}, parent=snibc}

\newglossaryentry{snii}{name=Type~II, text={SN~II}, description={Collapse of the core of a massive star - spectrum shows strong hydrogen line},first={Type~II supernova (SN~II)}, firstplural={Type~II supernovae (SNe~II)}, plural={SNe~II}, parent=sn}

\newglossaryentry{sniib}{name=Type~IIb, text={SN~IIb}, description={Spectrum shows hydrogen and helium lines},first={Type~IIb supernova (SN~IIb)}, firstplural={Type~IIb supernovae (SNe~IIb)}, plural={SNe~IIb}, parent=snii}

\newglossaryentry{sniip}{name=Type~II~Plateau (Type IIP), text={SN~IIP}, description={Lightcurve shows plateau},first={Type~IIP supernova (SN~IIP)}, firstplural={Type~II Plateau supernovae \citep[SNe~IIP;][]{1979A&A....72..287B}}, plural={SNe~IIP}, parent=snii}

\newglossaryentry{sniil}{name=SN~II~Linear, text={SN~IIL}, description={Lightcurve shows no plateau, but linear decline},first={Type~IIL supernova (SN~IIL)}, firstplural={Type~II~Linear supernovae \citep[SNe~IIL;][]{1990MNRAS.244..269S}}, plural={SNe~IIL}, parent=snii}

\newglossaryentry{sniin}{name=Type II narrow-lined (Type IIn), description={Spectrum shows narrow lines},first={Type~II~narrow-lined supernova (SN IIn)}, firstplural={Type~IIn supernovae (SNe~IIn)}, plural={SNe~IIn}, parent=snii}

\newglossaryentry{snr}{name=Remnant (SNR), text=SNR, description={Remnant left visible post-explosion}, first={supernova remnant (SNR)}, firstplural={supernova remnants (SNRs)}, parent=sn}

\newglossaryentry{dtd}{name=DTD,description={delay time distribution - expected supernova rate over time after a brief outburst of starformation},first={delay time distribution (DTD)}, firstplural={delay time distributions (DTDs)}, plural=DTDs}

\newglossaryentry{hvg}{name=HVG,description={high velocity gradient - Type Ia supernovae with a fast evolution of photospheric velocity},first={high velocity group (HVG)}, firstplural={high velocity groups (HVGs)}, plural=HVGs, parent=snia}

\newglossaryentry{lvg}{name=LVG,description={low velocity gradient - Type Ia supernovae with a slow evolution of photospheric velocity},first={low velocity group (LVG)}, firstplural={low velocity groups (LVGs)}, plural=LVGs, parent=snia}

\newglossaryentry{wd}{name=white dwarf (WD), text=WD, description={White Dwarf - extremely dense stellar remnant}, first={white dwarf (WD)}}
\newglossaryentry{onemgwd}{name= Oxygen/Neon (ONe), text={ONe-WD},description={Oxygen/Neon White Dwarf}, first={oxygen/neon White Dwarf (ONe-WD)}, parent=wd}
\newglossaryentry{cowd}{name=carbon/oxygen (CO), text={CO-WD}, description={carbon/oxygen white dwarf}, first={carbon/oxygen white dwarf (CO-WD)}, firstplural = {carbon/oxygen white dwarfs (CO-WDs)}, parent=wd}

\newglossaryentry{sds}{name=SD-Scenario,description={single-degenerate scenario (single white dwarf accreting from non-degenerate companion)}, first={single-degenerate scenario (SD-scenario)}}

\newglossaryentry{dds}{name=DD-Scenario, description={double degenerate scenario (merging of two white dwarfs)}, first={double-degenerate scenario (DD-scenario)}}

\newglossaryentry{sss}{name=SSS, text={supersoft \xray\ source}, description={supersoft \xray\ source - believed to be emitted by nuclear fusion on a white dwarf's surface}}%, first={supersoft \xray\ source (SSS)}, firstplural={supersoft \xray\ sources (SSS)}}

\newglossaryentry{amcvn}{name=AM CVn, description={AM Canum Venaticorum star \citep[white dwarf accreting hydrogen poor matter from a companion star; see ][]{2005ASPC..330...27N}}}

\newglossaryentry{rlof}{name=RLOF, description={Roche Lobe Overflow (see \citet{1971ARA&A...9..183P} for a more detailed description)}, first={Roche-lobe overflow (RLOF)}}

\newglossaryentry{mchan}{name={Chandrasekhar mass~}, text={Chandrasekhar~mass}, symbol={\ensuremath{M_\textrm{Chan}}}, plural={Chandrasekhar~masses}, description={Mass when the core of a star collapses due to insufficient degeneracy pressure - for a white dwarf $\approx1.38\,M_\odot$ see \citet{1931ApJ....74...81C}}, first={Chandrasekhar~mass \citep[$M_\textrm{Chan}=1.38\,M_\odot$;][]{1931ApJ....74...81C}}, sort=mchan}

\newglossaryentry{w7}{name={W7 model},description={W7 model \citep{1984ApJ...286..644N}},first = {W7 model \citep{1984ApJ...286..644N}}}

\newglossaryentry{ew}{name=Equivalent Width, text={EW}, description={width of a rectangle that has the same area as a spectral line when taken to zero flux}, first={equivalent width (EW)}, firstplural={equivalent widths (EWs)}}
\newglossaryentry{agb}{name=AGB,description={Asymptotic Giant Branch}, first={Asymptotic Giant Branch (AGB)}}
\newglossaryentry{cmb}{name=CMB,description={Cosmic Microwave Background}}
\newglossaryentry{csm}{name=CSM,description={Circumstellar Medium}, first={circumstellar medium (CSM)}}
\newglossaryentry{csi}{name=CSI,description={Circumstellar Interaction}, first={circumstellar interaction (CSI)}}
\newglossaryentry{ism}{name=ISM,description={Interstellar Medium}, first={interstellar medium (ISM)}}
\newglossaryentry{ige}{name=IGE,description={Iron Group Element}, first={iron group element (IGE)}, firstplural={iron group elements (IGEs)}}
\newglossaryentry{epm}{name=EPM,description={Expanding Photosphere Method \citep{1974ApJ...193...27K}}, first={Expanding Photosphere Method (EPM)}}
\newglossaryentry{aic}{name=AIC,description={Accretion Induced Collapse}, first={accretion induced collapse (AIC)}}
\newglossaryentry{ime}{name=IME,description={Intermediate Mass Element}, first={intermediate mass element (IME)}, firstplural={intermediate mass elements (IMEs)}}
\newglossaryentry{h0}{name=\ensuremath{H_0},description={Hubbles constant}}
\newglossaryentry{nse}{name=NSE,description={Nuclear Statistical Equilibrium}, first={nuclear statistical equilibrium (NSE)}}
\newglossaryentry{cdm}{name=CDM,description={Cold Dark Matter}}
\newglossaryentry{grb}{name=GRB,description={Gamma Ray Burst}, first={Gamma Ray Burst (GRB)}, firstplural={Gamma Ray Bursts (GRBs)}}
\newglossaryentry{xps}{name=XPS, description={X-ray point source}, first={X-ray point source (XPS)}, firstplural={X-ray point sources (XPS)}}
\newglossaryentry{donor}{name=donor,description={non-degenerate companion in the \gls{sds}}}
\newglossaryentry{mainsequence}{name=main sequence,description={main sequence star}}
\newglossaryentry{redgiant}{name=red giant,description={red giant star}}
\newglossaryentry{mlcs}{name=MLCS,description={Multicolor Light Curve Shape method \citep[MLCS;][]{1996ApJ...473...88R}}, first={Multicolor Light-Curve Shape method \citep[MLCS;][]{1996ApJ...473...88R}}}
\newglossaryentry{rsoph}{name=RS~Ophiuci ,description={white dwarf accreting from a red giant - assumed progenitor of the \gls{sds}}, sort=rsoph}
\newglossaryentry{usco}{name=U~Scorpii,description={white dwarf accreting from a main sequence star - assumed progenitor of the \gls{sds}}, sort=usco}
\newglossaryentry{rcw86}{name=RCW~86,description={supernova remnant sometimes associated with \sn{185}{}}, sort=rcw86}
\newglossaryentry{casa}{name=Cas~A,description={Cassiopeia A supernova remnant - probably a \gls{snib} event}}
\newglossaryentry{cepheid}{name=Cepheid,description={very luminous variable star with a strong luminosity period relationship}}
\newglossaryentry{urca}{name=Urca, text=\textit{Urca}, description={process predominatly contributing to cooling in stars. The \textit{Urca} process consists of alternating electron-capture and $\beta^{-}$ decay of two nuclei pairs.},sort=urca}
\newglossaryentry{alphacen}{name=Alpha Centauri,description={one of the brightest stars in the night sky and a close binary}}
\newglossaryentry{pcygni}{name={P Cygni}, text={P Cygni},description={a hypergiant luminous blue variable with strong winds. Often referred to as a description for their line profiles showing a emission peak at the rest wavelength of the line and a blue-shifted absorption trough.}}

\newglossaryentry{teff}{name={effective temperature~}, text={effective temperature}, symbol={\ensuremath{T_\textrm{eff}}}, description={Temperature of a blackbody emitting the same total energy}, sort=teff}

\newglossaryentry{logg}{name={surface gravity~}, text={surface gravity}, symbol={\ensuremath{\textrm{log}\,g}}, description={gravity at the surface of a star}, sort=logg}
\newglossaryentry{feh}{name={metallicity~}, text={metallicity}, symbol=\textrm{[Fe/H]},description={iron abundance relative to the sun}, sort=feh}

\newglossaryentry{texp}{name={time since explosion~}, text={time since explosion}, text={time since explosion}, symbol={\ensuremath{t_{\rm exp}}},description={time since explosion (measured in days)}, sort=texp, first={time since explosion (\ensuremath{t_{\rm exp}})}}

\newglossaryentry{lmc}{name=LMC,description={Large Magellanic Cloud}, first={Large Magellanic Cloud (LMC)}, sort=lmc}
\newglossaryentry{smc}{name=SMC,description={Small Magellanic Cloud}, sort=smc}
\newglossaryentry{z}{name=\ensuremath{z},description={redshift}, sort=z}

\newglossaryentry{stats.pdf}{name=PDF, description={Probability Density Function}, first={Probability Density Function}}

\makeglossaries

\usepackage{threeparttable}

\graphicspath{{./}{figures/}}

\submitjournal{ApJ}

\shorttitle{SN~2019ewu: A Peculiar Supernova from a New Sample of Young SN Ic Spectra}
\shortauthors{Williamson et al.}

\begin{document}

\title{SN~2019ewu: A Peculiar Supernova with Early Strong Carbon and Weak Oxygen Features from a New Sample of Young SN Ic Spectra}

\author[0000-0003-2544-4516]{Marc Williamson}
\affil{New York University}
\affil{TARDIS Collaboration}

\author[0000-0002-7941-5692]{Christian Vogl}
\affil{TARDIS Collaboration}
\affil{Max-Planck-Institut für Astrophysik}

\author[0000-0001-7132-0333]{Maryam Modjaz}
\affil{University of Virginia}

\author[0000-0002-0479-7235]{Wolfgang Kerzendorf}
\affil{TARDIS Collaboration}
\affil{Michigan State University}

\author[0000-0002-8310-0829]{Jaladh Singhal}
\affil{TARDIS Collaboration}

%GSP Co-authors
\author{Teresa Boland}
\affil{Department of Physics and Astronomy, Rutgers, the State University of New Jersey}

\author{Jamison Burke}
\affil{Las Cumbres Observatory, 6740 Cortona Dr, Suite 102, Goleta, CA 93117-5575, USA}

\author{Zhihao Chen}
\affil{Las Cumbres Observatory, 6740 Cortona Dr, Suite 102, Goleta, CA 93117-5575, USA}

\author[0000-0002-1125-9187]{Daichi Hiramatsu}
\affil{Las Cumbres Observatory, 6740 Cortona Dr, Suite 102, Goleta, CA 93117-5575, USA}

\author[0000-0002-1296-6887]{Llu\'is Galbany}
\affil{Institute of Space Sciences (ICE, CSIC), Campus UAB, Carrer de Can Magrans, s/n, E-08193 Barcelona, Spain}
\affil{Institut d’Estudis Espacials de Catalunya (IEEC), E-08034 Barcelona, Spain}

\author[0000-0003-0209-9246]{Estefania Padilla Gonzalez}
\affil{Las Cumbres Observatory, 6740 Cortona Dr, Suite 102, Goleta, CA 93117-5575, USA}
\affil{Department of Physics, University of California, Santa Barbara, CA 93106-9530, USA}

\author{D. Andrew Howell}
\affil{Las Cumbres Observatory, 6740 Cortona Dr, Suite 102, Goleta, CA 93117-5575, USA}
\affil{Department of Physics, University of California, Santa Barbara, CA 93106-9530, USA}

\author{Saurabh W. Jha}
\affil{Department of Physics and Astronomy, Rutgers, the State University of New Jersey}

\author{Lindsey A. Kwok}
\affil{Department of Physics and Astronomy, Rutgers, the State University of New Jersey}

\author{Curtis McCully}
\affil{Las Cumbres Observatory, 6740 Cortona Dr, Suite 102, Goleta, CA 93117-5575, USA}

\author{Megan Newsome}
\affil{Las Cumbres Observatory, 6740 Cortona Dr, Suite 102, Goleta, CA 93117-5575, USA}

\author[0000-0002-7472-1279]{Craig Pellegrino}
\affil{Las Cumbres Observatory, 6740 Cortona Dr, Suite 102, Goleta, CA 93117-5575, USA}

\author{Jeonghee Rho}
\affil{SETI Institute, 339 N. Bernardo Ave., Ste. 200, Mountain View, CA 94043; jrho@seti.org}

\author{Giacomo Terreran}
\affil{Las Cumbres Observatory, 6740 Cortona Dr, Suite 102, Goleta, CA 93117-5575, USA}
\affil{Department of Physics, University of California, Santa Barbara, CA 93106-9530, USA}

\author{Xiaofeng Wang}
\affil{Physics Department and Tsinghua Center for Astrophysics, Tsinghua University, Beijing, 100084, China}
\affil{Beijing Planetarium, Beiing Academy of Sciences and Technology, Beijing, 100044, China}

\begin{abstract}
With the advent of high cadence, all-sky automated surveys, supernovae (SNe) are now discovered closer than ever to their dates of explosion. However, young pre-maximum light follow-up spectra of Type Ic supernovae (SNe Ic), probably arising from the most stripped massive stars, remain rare despite their importance. In this paper we present a set of 49 optical spectra observed with the Las Cumbres Observatory through the Global Supernova Project for 6 SNe Ic, including a total of 17 pre-maximum spectra, of which 8 are observed more than a week before V-band maximum light. This dataset increases the total number of publicly available pre-maximum light SN Ic spectra by 25\% and we provide publicly available SNID templates that will significantly aid in the fast identification of young SNe Ic in the future. We present detailed analysis of these spectra, including Fe II 5169 velocity measurements, O I 7774 line strengths, and continuum shapes. We compare our results to published samples of stripped supernovae in the literature and find one SN in our sample that stands out. SN~2019ewu has a unique combination of features for a SN Ic: an extremely blue continuum, high absorption velocities, a P-cygni shaped feature almost 2 weeks before maximum light that TARDIS radiative transfer modeling attributes to C II rather than H$\alpha$, and weak or non-existent O I 7774 absorption feature until maximum light.

\end{abstract}

\keywords{Young supernovae, Core collapse supernovae, Supernovae Ic, Optical spectroscopy, Radiative transfer simulations, SN~2019ewu}

\section{Introduction} \label{sec:intro}

Stripped-Envelope (SE) supernovae (SNe) whose photospheric optical spectra lack both hydrogen and helium absorption features (SNe Ic; \citealt{filippenko1993type,modjaz19_review}) are the explosions of massive stars that have lost their outermost layers \citep{clocchiatti1997sn}, either through strong winds \citep{yoon2017towards}, binary interactions \citep{woosley2020ibc,dessart2020supernovae}, or both. The spectral diversity within the SN Ic class is well documented \citep{liu16,Williamson19} and indicative of a variety of different progenitor systems, mass-loss histories, and explosion mechanics contributing to SN Ic events.  For example, a subset of high velocity SNe Ic with broad lines (SNe Ic-bl) are the only type of SN observed in conjunction with gamma ray bursts \citep{modjaz_icbl}.  In the last two decades, a handful of SN Ic and SESNe sample studies have yielded valuable insights into this important supernova class.  For example, it has been shown that SNe Ic have stronger oxygen features than other SESN subclasses \citep{matheson2001optical, liu16, fremling_ptf_sesne}.  SNe Ic also have systematically higher velocities than their helium rich counterparts SNe Ib \citep{liu16,fremling_ptf_sesne}. While SNe Ic are supposedly helium free, there has been much debate over potential helium contribution in well observed SNe Ic (e.g. 1994I; \citealt{williamson_94i}, 2020oi; \citealt{rho_2020oi}), and a recent sample of near infrared (NIR) SN Ic spectra covering the unblended 2 micron line shows clear evidence of weak helium features in some SNe Ic \citep{shahbandeh_CSP_sesn}.  These sample studies use spectral datasets that mostly cover SN Ic evolution near and after the date of maximum brightness due to the difficulty of discovering SNe soon after explosion with low cadence surveys. The Berkeley sample of 888 SESN spectra collected over three decades by the Lick Observatory Supernova Search \citep[LOSS; ][]{shivvers2018berkeley} contains only 16 SN Ic spectra that were taken before maximum light. The Palomar Transient Factory (PTF) and intermediate PTF (iPTF) SESN dataset of 507 spectra \citep[][]{fremling_ptf_sesne} contains only 46 pre-maximum SN Ic spectra, most of which are not yet publicly available. A set of 156 SESN spectra released by the Public ESO Spectroscopic Survey of Transient Objects (PESSTO; \citealt{smartt_pessto}) contains only 16 pre-maximum SNe Ic spectra \citep{pessto_sesne}. The largest publicly available dataset of young SN Ic spectra contains 65 pre-maximum (relative to V-band) spectra from 18 SNe Ic \citep[hereafter the Modjaz Group Sample (MGS); ][]{modjaz2014optical,liu16, liu2017_slsne,Williamson19}. The MGS includes SESNe from the sample collected by the Harvard-Smithsonian Center
for Astrophysics (CfA; \citealt{modjaz_14cfa}), SNe Ic individually published by both the Berkeley and PTF/iPTF samples, as well as individual SNe Ic spectra published in the literature, up to 2019 \citep[][]{Williamson19}. 

The relative lack of young SN Ic spectra is a source of major uncertainty when it comes to connecting progenitor models to these explosions. Young SN Ic spectra taken only days after explosion (i.e. 1-2 weeks before V-band maximum light) give the most direct constraints on the outermost layers of the SN ejecta. At these phases, the photosphere (and therefore the line forming region above it) is located far out in the ejecta, and lines are less contaminated by lower velocity ejecta material. Therefore young SN Ic spectra may offer the possibility of detecting early weak signatures of hydrogen or helium before these features are lost due to line blending, cooling of the ejecta, decreasing density, or a combination of these effects as the supernova evolves. In addition, young SN Ic spectra offer the possibility of detecting narrow emission from flash ionization due to shock breakout (flash spectroscopy; \citealt{gal2014wolf,modjaz19_review,kochanek2019physics}). Young SN Ic spectra are also crucial for making confident classifications fast enough for other follow-up to be triggered by or before maximum light. This is particularly important for space-based ultraviolet (UV) spectroscopy using the Hubble Space Telescope (HST). SN Ic UV spectra at early times constrain the heavy metal content of the outer layers of the ejecta \citep[e.g. ][]{ashall2019grb} yielding information on mixing during the progenitor's evolution and explosion. In addition, similar to SNe Ia \citep[][]{mazzali00}, SNe Ic optical spectra are strongly influenced by high UV opacity due to heavy metal lines. Without templates for young SN Ic spectra, misclassifications can lead to inadvertent follow-up of the wrong SN type \citep[e.g. SN~2021yja; ][]{sergiy_21yja}. Young SNe Ic spectra in the nearby universe will prove necessary for identifying high redshift SNe Ic, where time dilation means observed spectra will be significantly younger in the SN rest frame. In addition, simulations in preparation for the Nancy Grace Roman Space Telescope (ROMAN) show that we can expect to find $\sim3000$ SNe Ia between $0.4\leq z \leq 1.7$ \citep{hounsell_roman}, and optical SN Ic spectra will help identify potential contaminants of young SNe Ia.

Two recently discovered SNe Ic (SN~2017ein; \citealt[][]{van2018sn17ein,kilpatrick2018potential}, SN~2020oi; \citealt{rho_2020oi,gagliano2022early}) have demonstrated the importance of young spectra for capturing peculiar behavior within the SN Ic class. SN~2017ein exhibited abnormally strong C II features at early times for an SN Ic and shared many other spectroscopic properties with SN~2007gr \citep[][]{valenti2008carbon}, the first so-called carbon-rich SN Ic \citep[][]{mazzali2010type_07gr}. Young spectra for both SN~2017ein and SN~2007gr show clear O I 7774 typical of SNe Ic \citep[][]{liu16,fremling_ptf_sesne}. Similarly, SN~2020oi also shows signs of early C II, but a spectrum observed just 3 days after explosion shows surprisingly weak O I 7774 for an SN Ic \citep[][]{gagliano2022early}. Carbon and oxygen features are particularly important for understanding SNe Ic because they can provide links to progenitor models. \citealt{Dessart12} have shown that models with higher ejecta mixing produce stronger and faster O I 7774 features, and recent stellar evolution simulations up to shock breakout have shown that a binary companion causes increased carbon yield in core collapse SNe \citep[][]{laplace_different_core,farmer_carbon}.

In this paper, we address the scarcity of young SN Ic spectra and present detailed modeling of the most peculiar object, SN~2019ewu. We present a publicly available dataset of 49 SN Ic spectra from 6 different events including 17 pre-maximum light spectra in Section \ref{sec:obs}. We discuss our analysis methodology in Section \ref{sec:methods} including how we correct for extinction, calculate spectral phases, characterize line velocity and strength, quantify continuum shape, and how we model SN~2019ewu. In Section \ref{sec:results} we present our results and compare the new SNe Ic in this paper to previous large samples from the literature (e.g. the MGS, PTF/iPTF Sample, Berkeley Sample). We show that SN~2019ewu exhibits unique early C II features for SNe Ic (both in strength and shape) as well as anomalously weak O I 7774. While a handful of SNe Ic in the literature have shown somewhate similar early carbon features, SN~2019ewu is the only SN Ic to exhibit such a weak O I 7774 absorption up to maximum light. We present models for SN~2019ewu using the radiative transfer code \gls{tardis} in Section \ref{sec:19ewu_model} and conclude in Section \ref{sec:summary}.

\section{Observations}

\label{sec:obs}
\begin{deluxetable*}{CCCCC}[t]
\tablecaption{Summary of SNe Spectra\label{tab:datatable}}
\tablecolumns{5}
\tablenum{1}
\tablewidth{0pt}

\startdata
        &        &          &        &    \\     
\textnormal{\textbf{SN Name}}& \textnormal{\textbf{SN Type}\tablenotemark{a}}&\textnormal{\textbf{z}}&$t_{V_{max}}$\textbf{(MJD)}&\textnormal{\textbf{Phases}\tablenotemark{b}} \\\hline
\textnormal{SN2019cda}& \textnormal{Ic}&0.027652&58575.9&\textnormal{-7.3, -6.5, 0.5, 7.3}\\
\textnormal{SN2019ewu}& \textnormal{Ic}&0.033&58625.4&\textnormal{-11.6, -4.9, 0.0, 9.7, 15.5, 28.0}\\
\textnormal{SN2019yz}& \textnormal{Ic}&0.00639&58516.2&\textnormal{-8.6, -7.6, 28.1, 38.1, 41.5, 46.0, 54.0, 61.9, 69.9, 82.7, 93.7}\\
\textnormal{SN2020akf}& \textnormal{Ic}&0.012&58883.8&\textnormal{-13.1, -12.2, -10.9, -4.4, -0.4, 27.2, 44.9,}\\
& & & &\textnormal{57.9, 65.8, 73.6, 83.5, 94.4, 109.2, 125.9} \\
% 
%\textnormal{SN2020oi}& \textnormal{Ic}&&\textnormal{todo}& &RHO\\
%
\textnormal{SN2021mxx}& \textnormal{Ic}&0.00965&59365.1&\textnormal{-7.6, -6.5, -4.7, -3.6, -0.7, 7.2, 14.3, 24.2, 33.1, 43.9, 61.7}\\
\textnormal{SN2021do}& \textnormal{Ic}&0.009346&59227.2&\textnormal{-0.7, 3.2, 27.9}\\
\enddata

\tablenotetext{a}{SNe are classified by running \gls{snid} with the most updated SESNe template library \citep[][]{liu16,modjaz_icbl,Williamson19}.}

\tablenotetext{b}{Phases are rounded to the nearest tenth and are in rest frame, relative to the date of V-band maximum. The date of peak in the available bands is fit with the monte carlo method of \citealt{bianco_vband_conv} and if V-band data is not available, the date of peak is converted to V-band using the empirical relations from \citep{bianco_vband_conv}. SNEX V-band photometry is used for SN~2019ewu and SN~2020akf, while LASAIR \citep{lasair} $r$-band photometry is used for the remaining objects.}
\end{deluxetable*}
We present a dataset of 49 spectra from 6 different SNe Ic obtained through the \gls{lcogt} collaboration between January 2019 and July 2021 as part of the Global Supernova Project (GSP). As part of the selection criteria, we choose SNe Ic with spectral time series that include at least one pre-maximum light spectrum (relative to V-band). The dataset presented here includes a total of 17 pre-maximum SN Ic spectra which is an increase of 25\% to the number of publicly available young SNe Ic spectra. Table \ref{tab:datatable} shows a summary of the dataset.

All 6 SNe presented here were eventually classified as SNe Ic, but some showed similarities to other SN types at early times. SN~2019cda was discovered \citep[][]{19cda_disc} by the Italian Supernovae Search Project (ISSP) and initially classified as type SN Ic using SuperFit \citep[][]{19cda_class}. SN~2019ewu was discovered \citep[][]{2019ewu_tns_discovery} by the ATLAS group and initially classified as type SN II due to a blue continuum and the appearance of a broad feature that was interpreted as H$\alpha$ \citep[][]{19ewu_class}. SN~2019yz was discovered \citep[][]{19yz_disc} by ZTF and initially classified as type SN Ic using SuperFit \citep[][]{19yz_class}. SN~2020akf was discovered \citep[][]{20akf_disc} by the ATLAS group and initially classified as type SN Ia using SuperFit \citep[][]{20akf_class}. SN~2021mxx was discovered \citep[][]{21mxx_disc} by ZTF and initially classified as type SN Ia by SuperFit \citep[][]{21mxx_class}. Finally, SN~2021do was discovered \citep[][]{21do_disc} by ZTF and initially classified as type SN Ic \citep[][]{21do_class}. Here we reclassify SN 2019ewu, SN 2020akf, and SN 2021mxx as SNe Ic based on their spectra at maximum light using \gls{snid} and the MGS template libraries. We note that half of the young SNe Ic in this sample were initially misclassified most likely due to lack of searly template spectra in the libraries of codes like \gls{snid}.

Table \ref{tab:obslog} details the observation log for each spectrum included. Optical spectra were taken using the FLOYDS spectrographs on the 2-meter telescopes at Siding Spring Observatory (COJ 2m) and Haleakala (OGG 2m). FLOYDS spectra cover the wavelength range from 3500--10000$\AA$ at a resolution varying between $400\leq R \leq 700$. We include one spectrum for SN~2020akf that was shared with the GSP and was observed by the Beijing Faint Object Spectrograph and Camera (BFOSC) using a 2.16 meter telescope (2.16BAO) covering a slightly narrower optical wavelength range at similar resolution of $500\leq R \leq2000$ \citep{fan2016xinglong}. 

\begin{deluxetable*}{LCCCCCC}[h!]
\tablecaption{Spectra Observation Log\label{tab:obslog}}
\tablecolumns{7}
\tablenum{2}
\tablewidth{0pt}
\startdata
&&&&&&\\
%\textnormal{\textbf{SN (phase)}}&\textnormal{\textbf{Telescope}}&\textnormal{\textbf{Instrument}}&\textnormal{\textbf{Exposure}}&\textnormal{\textbf{Slit}}&\textnormal{\textbf{Acq. Rad.}}&\textnormal{\textbf{Airmass}}\\
\textnormal{\textbf{SN }} (t_{V_{max}}\textnormal{ days})&\textnormal{\textbf{Telescope}}&\textnormal{\textbf{Instrument}}&\textnormal{\textbf{Exposure (s)}}&\textnormal{\textbf{Slit (arcsec)}}&\textnormal{\textbf{Airmass}}\\
\hline
%-7.4, -6.6, 0.5, 7.5
\textnormal{2019cda (-7.3)}&\textnormal{OGG 2m}&\textnormal{FLOYDS}&3600&2''&1.25\\
\textnormal{2019cda (-6.5)}&\textnormal{OGG 2m}&\textnormal{FLOYDS}&3600&2''&1.26\\
\textnormal{2019cda (0.5)}&\textnormal{OGG 2m}&\textnormal{FLOYDS}&3600&2''&1.28\\
\textnormal{2019cda (7.3)}&\textnormal{OGG 2m}&\textnormal{FLOYDS}&3600&2''&1.25\\
\hline
%
%-12.0, -5.1, 0.0, 10.0, 16.0, 28.9
\textnormal{2019ewu (-11.6)}&\textnormal{OGG 2m}&\textnormal{FLOYDS}&3600&2''&1.38\\
\textnormal{2019ewu (-4.9)}&\textnormal{OGG 2m}&\textnormal{FLOYDS}&3600&2''&1.21\\
\textnormal{2019ewu (0.0)}&\textnormal{OGG 2m}&\textnormal{FLOYDS}&3600&2''&1.46\\
\textnormal{2019ewu (9.7)}&\textnormal{OGG 2m}&\textnormal{FLOYDS}&3600&2''&1.41\\
\textnormal{2019ewu (15.5)}&\textnormal{OGG 2m}&\textnormal{FLOYDS}&3600&2''&1.42\\
\textnormal{2019ewu (28.0)}&\textnormal{OGG 2m}&\textnormal{FLOYDS}&3600&2''&1.31\\
\hline
%
%-8.6, -7.6, 28.3, 38.3, 46.4, 54.4, 62.3, 70.4, 83.3, 94.4
\textnormal{2019yz (-8.6)}&\textnormal{OGG 2m}&\textnormal{FLOYDS}&2700&2''&1.74\\
\textnormal{2019yz (-7.6)}&\textnormal{OGG 2m}&\textnormal{FLOYDS}&2700&2''&1.52\\
\textnormal{2019yz (28.1)}&\textnormal{OGG 2m}&\textnormal{FLOYDS}&2700&2''&1.50\\
\textnormal{2019yz (38.1)}&\textnormal{OGG 2m}&\textnormal{FLOYDS}&2700&2''&1.14\\
\textnormal{2019yz (41.5)}&\textnormal{SALT}&\textnormal{RSS}&2194&1.5''&1.26\\
\textnormal{2019yz (46.0)}&\textnormal{OGG 2m}&\textnormal{FLOYDS}&3600&2''&1.08\\
\textnormal{2019yz (54.0)}&\textnormal{OGG 2m}&\textnormal{FLOYDS}&3600&2''&1.08\\
\textnormal{2019yz (61.9)}&\textnormal{OGG 2m}&\textnormal{FLOYDS}&3600&2''&1.09\\
\textnormal{2019yz (69.9)}&\textnormal{COJ 2m}&\textnormal{FLOYDS}&3600&2''&1.40\\
\textnormal{2019yz (82.7)}&\textnormal{OGG 2m}&\textnormal{FLOYDS}&3600&2''&1.09\\
\textnormal{2019yz (93.7)}&\textnormal{COJ 2m}&\textnormal{FLOYDS}&3600&2''&1.20\\
\hline
%
%-13.2, -12.4, -11.1, -4.5, -0.4, 27.6, 45.4, 58.6, 66.6, 74.5, 84.5, 95.5, 110.5, 127.5
\textnormal{2020akf (-13.1)}&\textnormal{OGG 2m}&\textnormal{FLOYDS}&3600&2''&1.35\\
\textnormal{2020akf (-12.2)}&\textnormal{OGG 2m}&\textnormal{FLOYDS}&3599&2''&1.06\\
\textnormal{2020akf (-10.9)}&\textnormal{2.16/BAO}&\textnormal{BFOSC}&3600&2.3''&1.01\\
\textnormal{2020akf (-4.4)}&\textnormal{OGG 2m}&\textnormal{FLOYDS}&2700&2''&1.36\\
\textnormal{2020akf (-0.4)}&\textnormal{OGG 2m}&\textnormal{FLOYDS}&2700&2''&1.17\\
\textnormal{2020akf (27.2)}&\textnormal{OGG 2m}&\textnormal{FLOYDS}&2700&2''&1.06\\
\textnormal{2020akf (44.9)}&\textnormal{OGG 2m}&\textnormal{FLOYDS}&3600&2''&1.14\\
\textnormal{2020akf (57.9)}&\textnormal{OGG 2m}&\textnormal{FLOYDS}&2700&2''&1.21\\
\textnormal{2020akf (65.8)}&\textnormal{OGG 2m}&\textnormal{FLOYDS}&2699&2''&1.42\\
\textnormal{2020akf (73.6)}&\textnormal{OGG 2m}&\textnormal{FLOYDS}&2700&2''&1.14\\
\textnormal{2020akf (83.5)}&\textnormal{OGG 2m}&\textnormal{FLOYDS}&2700&2''&1.17\\
\textnormal{2020akf (94.4)}&\textnormal{OGG 2m}&\textnormal{FLOYDS}&3599&2''&1.39\\
\textnormal{2020akf (109.2)}&\textnormal{OGG 2m}&\textnormal{FLOYDS}&3600&2''&1.47\\
\textnormal{2020akf (125.9)}&\textnormal{OGG 2m}&\textnormal{FLOYDS}&3600&2''&1.73\\
\hline
%
%-7.7, -6.6, -4.7, -3.7, -0.7, 7.3, 14.4, 24.4, 33.4, 44.3, 62.3
\textnormal{2021mxx (-7.6)}&\textnormal{OGG 2m}&\textnormal{FLOYDS}&2700&2''&1.38\\
\textnormal{2021mxx (-6.5)}&\textnormal{OGG 2m}&\textnormal{FLOYDS}&2700&2''&1.07\\
\textnormal{2021mxx (-4.7)}&\textnormal{OGG 2m}&\textnormal{FLOYDS}&2699&2''&1.44\\
\textnormal{2021mxx (-3.6)}&\textnormal{COJ 2m}&\textnormal{FLOYDS}&2700&2''&1.07\\
\textnormal{2021mxx (-0.7)}&\textnormal{OGG 2m}&\textnormal{FLOYDS}&2700&2''&1.36\\
\textnormal{2021mxx (7.2)}&\textnormal{COJ 2m}&\textnormal{FLOYDS}&1800&2''&1.34\\
\textnormal{2021mxx (14.3)}&\textnormal{COJ 2m}&\textnormal{FLOYDS}&1800&2''&1.05\\
\textnormal{2021mxx (24.2)}&\textnormal{COJ 2m}&\textnormal{FLOYDS}&1800&2''&1.15\\
\textnormal{2021mxx (33.1)}&\textnormal{COJ 2m}&\textnormal{FLOYDS}&2700&2''&1.28\\
\textnormal{2021mxx (43.9)}&\textnormal{COJ 2m}&\textnormal{FLOYDS}&2700&2''&1.08\\
\textnormal{2021mxx (61.7)}&\textnormal{COJ 2m}&\textnormal{FLOYDS}&2699&2''&1.08\\
\hline
%
%-0.7, 3.3, 28.2
\textnormal{2021do (-0.7)}&\textnormal{OGG 2m}&\textnormal{FLOYDS}&2699&2''&1.71\\
\textnormal{2021do (3.2)}&\textnormal{OGG 2m}&\textnormal{FLOYDS}&2700&2''&1.69\\
\textnormal{2021do (27.9)}&\textnormal{OGG 2m}&\textnormal{FLOYDS}&3600&2''&1.70\\
\enddata
\end{deluxetable*}

\section{Methodology}
\label{sec:methods}
In this section we discuss the choices made to process and analyze the SNe Ic spectra in our dataset. This includes extinction correction, classification, phase calculation, and spectral feature and continuum shape characterization. The spectra are extracted from raw images using the FLOYDS Pipeline\footnote{\url{https://lco.global/documentation/data/floyds-pipeline/}} which includes cosmic ray cleaning, trace extraction, and wavelength and flux calibration.

\subsection{Extinction Correction}
\label{sec:ext}
\begin{deluxetable}{LCCC}[t!]
\tablecaption{Extinction\label{tab:ext_tab}}
\tablecolumns{4}
\tablenum{3}
\tablewidth{0pt}

\startdata
        &        &          &            \\     
\textnormal{\textbf{SN}}& \textnormal{\textbf{RA}}&\textnormal{\textbf{DEC}}&\textbf{E(B-V)\tablenotemark{(a)}}$_{MW}$ \\
&(\textnormal{h m s})&($^{\circ}$\textnormal{ ' ''})&(\textnormal{mag})\\
\hline
\textnormal{19cda}&\textnormal{10:44:49.84} &\textnormal{+06:36:04.0}&0.0269\\
\textnormal{19ewu}&\textnormal{13:17:12.25}&\textnormal{+53:54:57.4}&0.0205 \\
\textnormal{19yz}\tablenotemark{(b)}&\textnormal{15:41:57.30}&\textnormal{+00:42:39.4}&0.0982 \\
\textnormal{20akf}&\textnormal{09:28:39.62}&\textnormal{+38:33:47.1}&0.0135 \\
% 
%\textnormal{SN2020oi}& \textnormal{Ic}&&\textnormal{todo}& &RHO\\
%
\textnormal{21mxx}&\textnormal{18:56:51.27}&\textnormal{+36:37:20.4}&0.0822 \\
\textnormal{21do}\tablenotemark{(b)}&\textnormal{10:16:56.67}&\textnormal{+73:23:51.3}&0.0207 \\
\enddata
\tablenotetext{a}{Milky Way extinction is calculated using \cite{schlafly2011measuring}.}
\tablenotetext{b}{Host galaxy extinction is evident from the presence of a narrow Na I D line at the host galaxy redshift. In this case, host extinction is calculated using Equation 1 from \cite{taubenberger2006sn}. See discussion in Section \ref{sec:ext}.}
\end{deluxetable}
For each SN, we calculate Milky Way (MW) extinction using \citealt{schlafly2011measuring} via the NASA/IPAC calculator\footnote{\url{https://irsa.ipac.caltech.edu/applications/DUST/}}. Only SN~2019yz and SN~2021do show narrow Na I D absorption features (at rest) at their host galaxy redshifts, so for these two SNe we calculate host galaxy extinction following the procedure in \citealt{taubenberger2006sn}. We find a host extinction of $E(B-V)=0.13$ mag for SN~2019yz from a narrow Na I D line with equivalent width 0.81$\AA$ and extinction of $E(B-V)=0.52$ mag for SN~2021do from a narrow Na I D line with equivalent width 3.25$\AA$. Extinction corrections are made assuming $R_{V}=3.1$ and a \citealt{fitzpatrick1999correcting} reddening law. These MW and host galaxy extinction correction choices are made specifically for consistency with the methodology used by \citealt{barbarino2021type} to analyze the SNe Ic from the Palomar Transient Factory \citep[PTF,][]{rau2009exploring,law2009palomar}. Table \ref{tab:ext_tab} shows the positions of the SNe in our dataset on the sky and their associated MW extinction correction.

\subsection{Spectrum Phases and Peak Calculation}
\label{sec:phases}

Each SN in our dataset has observed photometry that covers the lightcurve peak in at least one band, so spectral phases are calculated relative to the date of maximum. Specifically, the date of maximum is calculated for the V-band either directly or using the conversion factor for the appropriate band \citep{bianco_vband_conv} for consistency with the analysis of the MGS, one of the largest publicly available spectral datasets of SNe Ic (\citealt{liu16,modjaz_icbl,Williamson19}). The date of maximum light is calculated using a Monte Carlo process to fit a quadratic around the lightcurve peak \citep{bianco_vband_conv}. \gls{lcogt} V-band photometry is used for SN~2019ewu and SN~2020akf and \gls{lasair} r-band difference photometry is used for the remaining SNe and the date of maximum converted to V-band. Table \ref{tab:datatable} shows the dates of maximum in the V-band ($t_{V_{max}}$) calculated in MJD for each of the SNe in our dataset.

\subsection{Spectral Classification}
\label{sec:snid}
We use the Supernova Identification code (SNID; \citealt{snid}) to classify the spectra in our dataset. \gls{snid} works by cross correlating the newly observed spectra in our dataset with a library of previously observed and typed SN spectral templates. In order to obtain robust classifications from \gls{snid}, it is imperative to use the most updated template libraries. In this case, we use the stripped supernova templates from the MGS \citep{liu16, modjaz_icbl, Williamson19}. For each SN in our sample, we run \gls{snid} on the spectrum closest to the date of maximum (i.e. $t_{V_{max}}=0$ days) to produce a robust classification. In addition, we run \gls{snid} on the earliest spectrum for each SN in order to check for consistency with initial classification attempts and to identify SNe with uncertain or evolving early types. We focus discussion on the three SNe in this sample that were originally classified as types other than SN Ic (SN~2019ewu, SN~2020akf, SN~2021mxx).

\subsection{Line Strengths and Velocities}
\label{sec:v_pew}
Line strengths and velocities are important quantitative tools for understanding SNe spectra in the context of previously observed SNe samples and for understanding SN dynamics and their progenitors. In order to include uncertainties in our reported line velocities and strengths, we first calculate uncertainty arrays for each SN spectrum in our dataset using the Fourier smoothing method developed by \citealt{liu16}. We use these uncertainty arrays instead of the uncertainty arrays directly from the reduction for consistent comparison of the uncertainties of our line velocities and strengths to those from the MGS. It has been shown that the Fourier-derived uncertainty arrays do not introduce systematic uncertainties \citep[see Fig. 17; ][]{liu16}. We calculate uncertainties for line velocities and strengths by resampling the spectra in our dataset hundreds of times using the uncertainty arrays to generate Gaussian noise in each wavelength bin. Line velocities and strengths are then calculated for each resampled spectrum to characterize the uncertainty in each calculation.

Line velocities are found by fitting a quadratic around the absorption minimum to identify a wavelength shift relative to the rest wavelength of a given line \citep[e.g. ][]{blondin2006using,liu16}. This wavelength shift is converted to a velocity using the relativistic doppler shift. In this work we focus on the Fe II 5169 line because it has previously been used as a tracer of the photospheric velocity in stripped envelope SNe \citep{liu16,modjaz_icbl,barbarino2021type}. We note that robust line velocities strongly depend on confident line identification, which can be difficult for core collapse SNe, which can exhibit high degrees of line blending due to high ejecta velocities, especially at early times. For this reason we restrict ourselves to calculating the Fe II 5169 line velocities to spectra where the characteristic 'W' Fe doublet feature is clearly visible.

We use the pseudo-equivalent width (pEW; \citealt{blondin2011spectra,silverman2012berkeley,liu16}) to quantify line strengths and particularly focus on the O I 7774 line for our SN Ic dataset. The O I 7774 line is relatively easy to identify (i.e. unblended), appears even in very early SN Ic spectra, and has been reported in previous SN Ic sample studies \citep[][]{matheson2001optical,liu16,barbarino2021type}. Furthermore, oxygen (the most abundant heavy element in the universe) may be tied to the progenitor structure. For example, some simulations have shown that lower mass progenitors produce smaller oxygen buffer regions between the inner and outer layers of SN ejecta, affecting non-thermal excitation that is crucial for SESNe \citep{Dessart12}. In addition, progenitors with similar core masses at time of collapse from binaries versus single star systems show systematic differences in carbon and oxygen structure \citep[][]{laplace_different_core}. To calculate the O I 7774 pEW, we follow the methodology of \citealt{liu16}, which is summarized below. Local maxima on either side of the O I 7774 absorption minimum are identified, and the local continuum is defined as a line connecting those points. The pEW is calculated using Equation 1 from \citealt{liu16}.

\subsection{Characterizing Young SN Ic Continuua}
\label{sec:blue}

In order to quantify the diverse continuum behavior exhibited by the SNe Ic in our dataset (particularly at early times), we calculate color evolution directly from the spectra using the ST Mag system\footnote{\url{https://hst-docs.stsci.edu/acsdhb}} used by HST. The ST Mag system is defined such that an object with constant flux per unit wavelength has zero color, so it is easily interpretable. We use the standard Bessell bands\footnote{\url{http://svo2.cab.inta-csic.es/theory/fps/index.php}} \citep[][]{bessell_bands} and the open-source package WSYNPHOT\footnote{\url{https://github.com/starkit/wsynphot}} to calculate magnitudes and colors directly from spectra. It has been shown that calculating colors directly from SESN spectra without absolute calibration to photometry introduces only minor uncertainty of $\sim0.15$ mag for phases before 20 days post maximum light \citep[see Fig. 10; ][]{modjaz_14cfa}.
\begin{figure*}[t]
 \centering
 \includegraphics[scale=0.30]{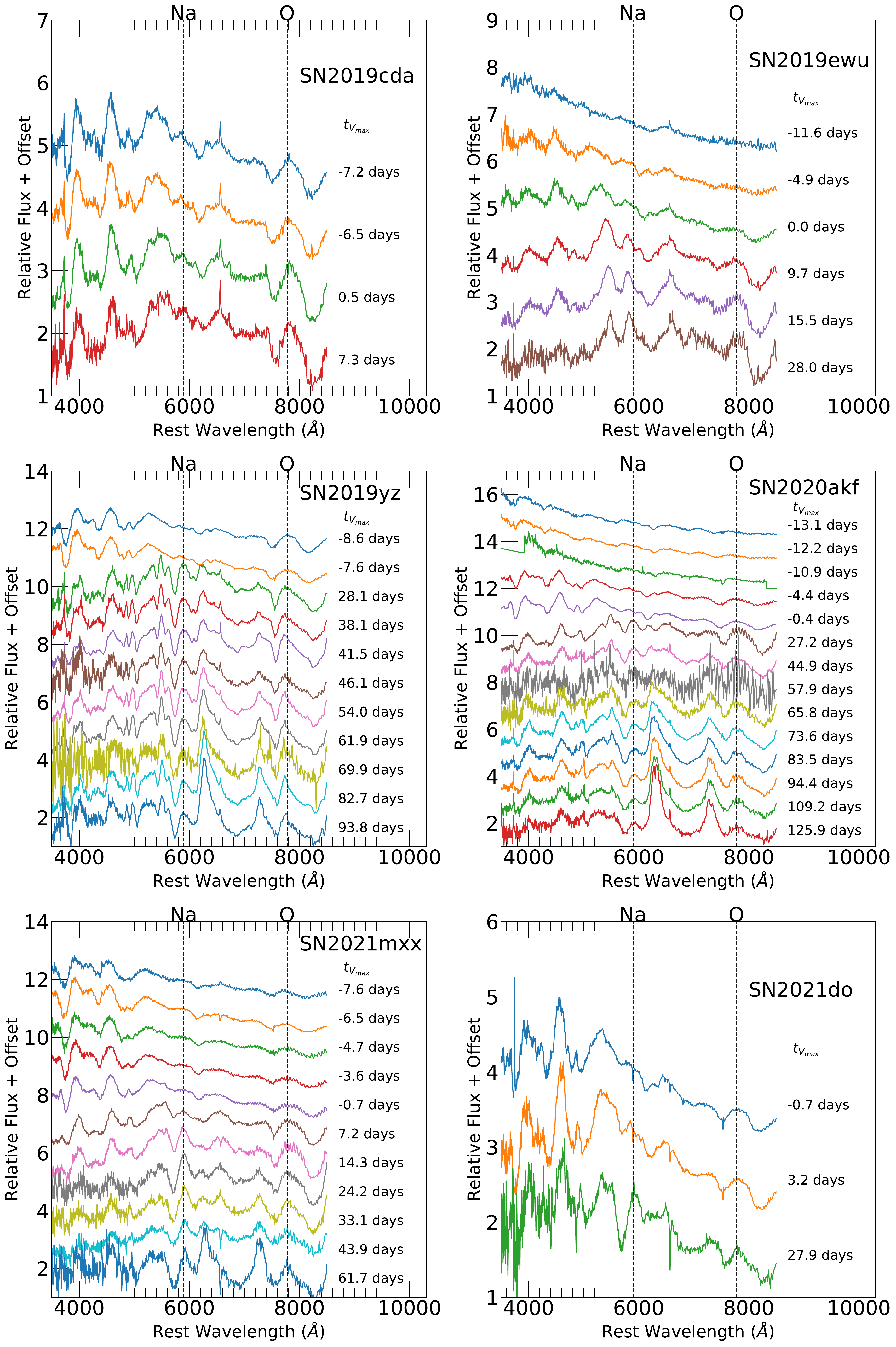}
 
    \caption{De-reddened (by both MW and host extinction when appropriate) spectral time series for the 6 SNe Ic analyzed in this work. Dashed vertical lines mark the Na 5896 and O 7774 lines, and phases are relative to the date of V-band maximum. All SNe with the exception of SN~2021do were discovered early enough to obtain spectra more than 1 week prior to maximum light. We note that SN~2019ewu and SN~2020akf have blue, featureless spectra at very early times for SNe Ic, which may be related to their slow-developing and weak O 7774 feature compared to the other SNe.}
     \label{fig:waterfalls}
      \vspace{0.0cm}

\end{figure*}
\subsection{Spectral Modeling with TARDIS}
\label{sec:tardis}

In order to robustly identify lines in blended spectra, radiative transfer simulations are required. \gls{tardis} is a fast, 1D, open-source radiative transfer code for modeling supernova spectra. \gls{tardis} has been used to model a wide variety of SN types, including SNe Ic. In particular, \gls{tardis} includes a non-local thermodynamic equilibrium (NLTE) approximation for the ionization and excitation treatment of Helium \citep[][]{boyle_tardis} which is important for modeling stripped envelope SNe. In this paper, we use two different versions of \gls{tardis} to model the earliest spectrum of SN~2019ewu, which contains an absorption feature that could potentially be attributed to Hydrogen (see Section \ref{sec:19ewu}). The public\footnote{\url{https://github.com/tardis-sn/tardis}} version of \gls{tardis} (v2022.06.19) currently does not include the continuum processes necessary for a accurate treatment of H (work is ongoing to incorporate the physics from \citealt{vogl_dil} which will include continuum processes in the public \gls{tardis}). We therefore use the public \gls{tardis} in order to establish a baseline model without H and to attempt to explain the SN~2019ewu absorption feature using C or Si. Our models assume a powerlaw density profile $\rho = \rho_{0}(v/v_{0})^{n}$ where the parameters are chosen to be consistent with the hydrodynamical simulations of an exploded stripped star \citep[][]{iwamoto94} that have been used for radiative transfer simulations of SN Ic 1994I \citep{hachinger94Imodel,williamson_94i}. We keep the exponent $n=-6.78$ fixed (fit from the outer layers of the hydrodynamical simulation) but vary the normalization parameter $\rho_{0}$ to experiment with different ejecta masses. We assume a date of explosion on May 07, 2019 according to the last non-detection \citep[][]{2019ewu_tns_discovery} because \gls{tardis} requires a parameter for the time since explosion. We note that due to the assumption of homologous expansion there is a degeneracy between the time since explosion and density profile parameters. Our primary focus with the models therefore is to understand the elements causing line formation in the ejecta.

We use the non-public version of \gls{tardis} presented in \citealt{vogl_dil}, which includes continuum processes and has been used to model H-rich SNe IIP \citep[][]{vogl_IIP_ML,sergiy_21yja} to modify the baseline model by adding H to the ejecta. We do not use the consistent thermal balancing implemented by \citealt{vogl_IIP_ML} since it neglects some heating and cooling processes for species other than H and He, which might make it unreliable for a hydrogen-poor SN. We instead follow the standard coupling of electron temperature to radiative temperature $T_{e}=0.9T_{rad}$ \citep[][]{mazzali1993application}. The non-public version of \gls{tardis} also includes an NLTE treatment for the ionization and excitation states of H.

\begin{figure}[t]
 \centering
 \includegraphics[scale=0.37]{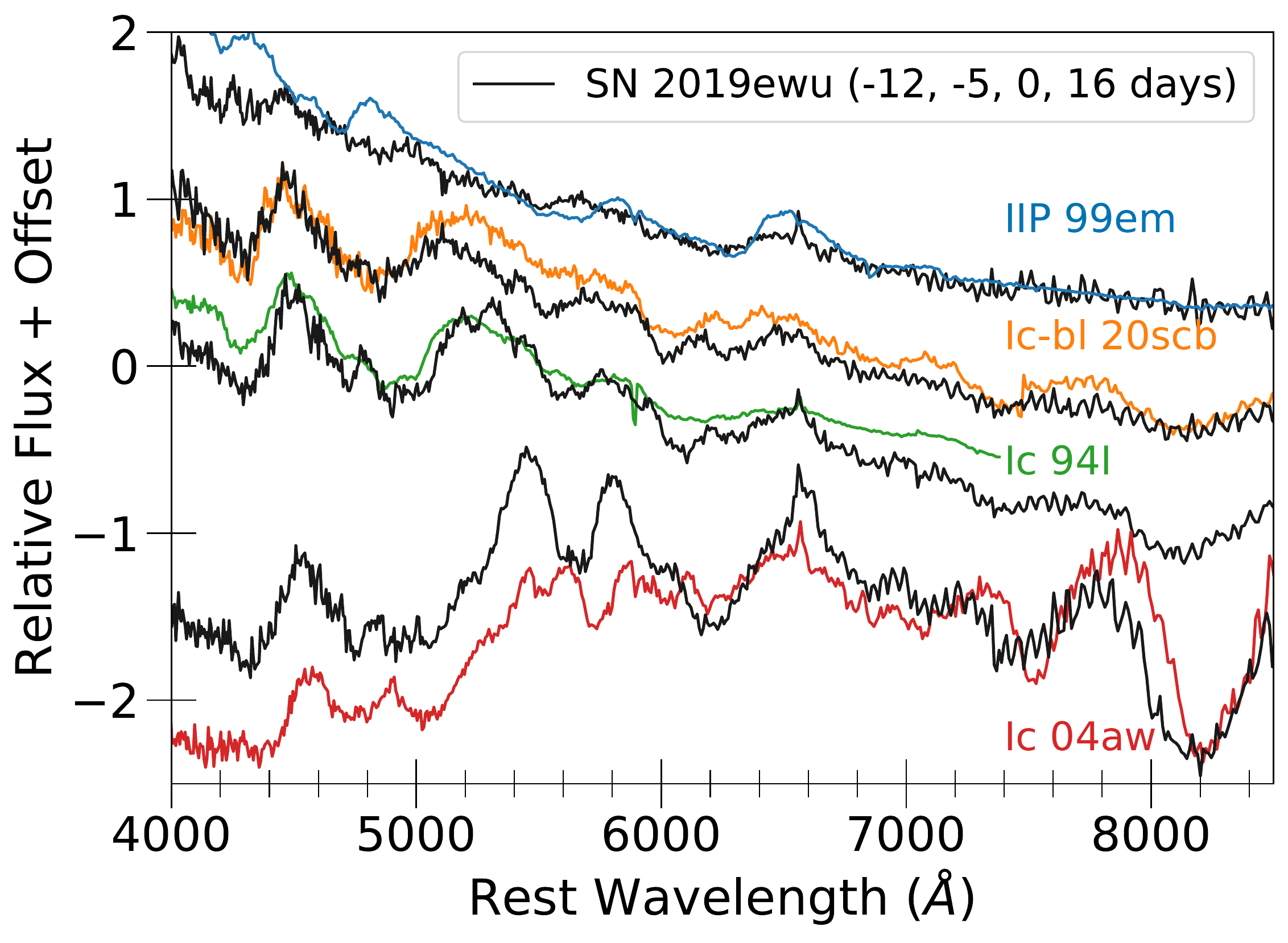}
 
    \caption{A subset of the SN~2019ewu spectral timeseries (black) at $-$12, $-$5, 0, and 16 days relative to V-band maximum. Colored spectra show particularly good spectral matches identified by SNID at each phase. SN~2019ewu initially matches with SN II before developing broad lines typical of SN Ic-bl, before settling as an SN Ic.}
     \label{fig:19ewu_types}
\end{figure}

%%%%%%%%%%%%waterfall%%%%%%%

\section{Results}
\label{sec:results}

Figure \ref{fig:waterfalls} shows the spectral time series of the SNe Ic in our dataset. The spectra shown are corrected for extinction as discussed in Section \ref{sec:ext} and each spectrum is labeled with the phase relative to V-band maximum as discussed in Section \ref{sec:phases}. The O I 7774 line is marked with a dashed vertical line to highlight the diversity in behavior observed in this wavelength region among the SNe. For example, SN~2019ewu shows no evidence (or very weak evidence) for an O I 7774 feature pre-maximum while SN~2019yz shows strong evidence for O I 7774 even more than a week before V-band maximum. We also mark the Na I D line (5893$\AA$) with a dashed vertical line to highlight the major contaminant for He I 5876 discussed for SNe Ic.

\begin{figure}[t]
 \centering
 \includegraphics[scale=0.35]{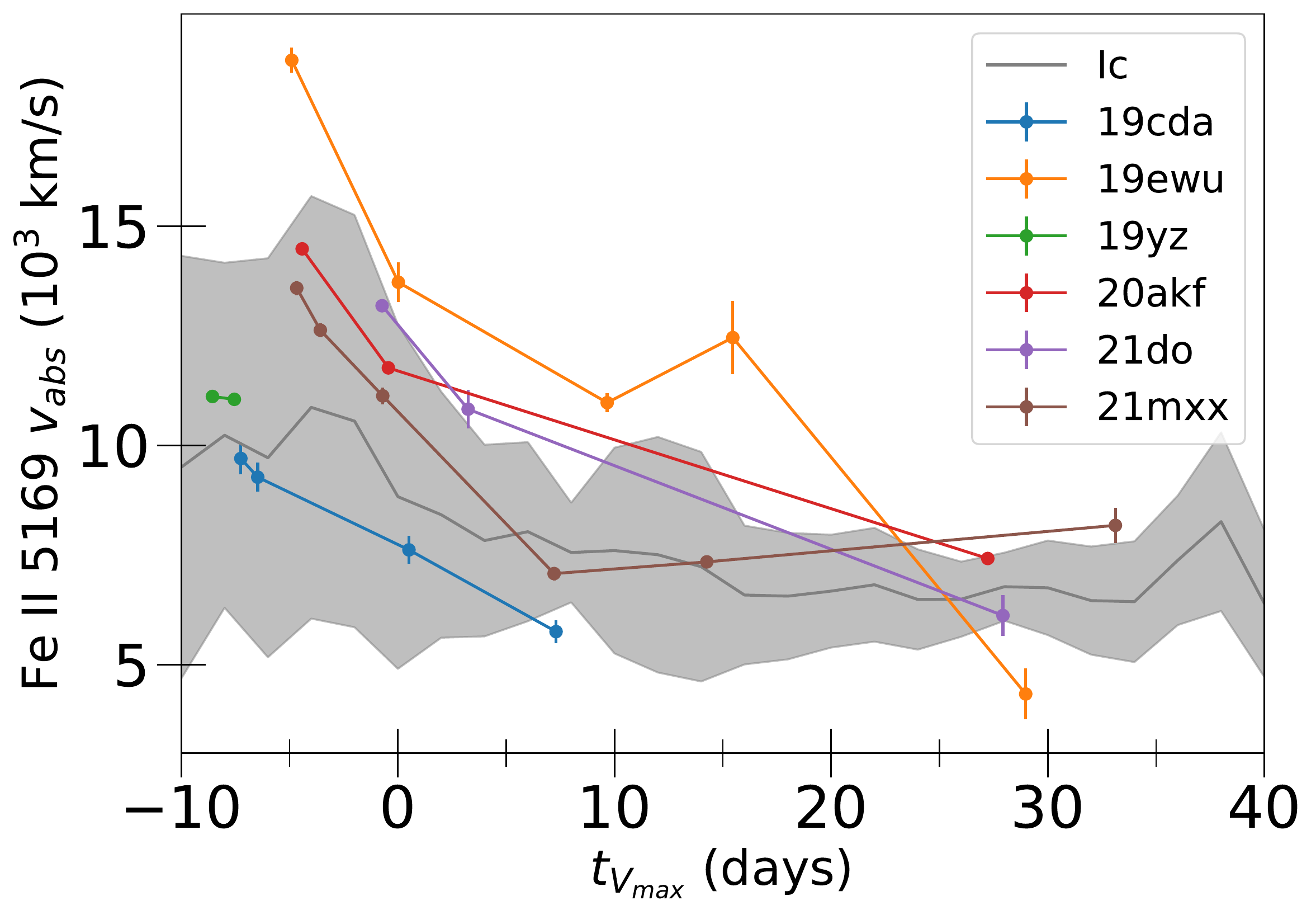}
 
    \caption{Fe II 5169 absorption velocities measured as a function of $t_{V_{max}}$ for the GSP SN Ic sample (colored lines) compared to the mean $\pm 1$ standard deviation velocity of the SN Ic sample from \citealt{liu16} shown as the grey band. Velocities are calculated only when Fe II 5169 can be clearly identified from the Fe doublet. A clear identification of Fe II 5169 cannot be made for SN~2019yz after early times due to contamination from an emergent feature.}
     \label{fig:fe_vel}

\end{figure}

\begin{figure*}[t]
 \centering
 \includegraphics[scale=0.35]{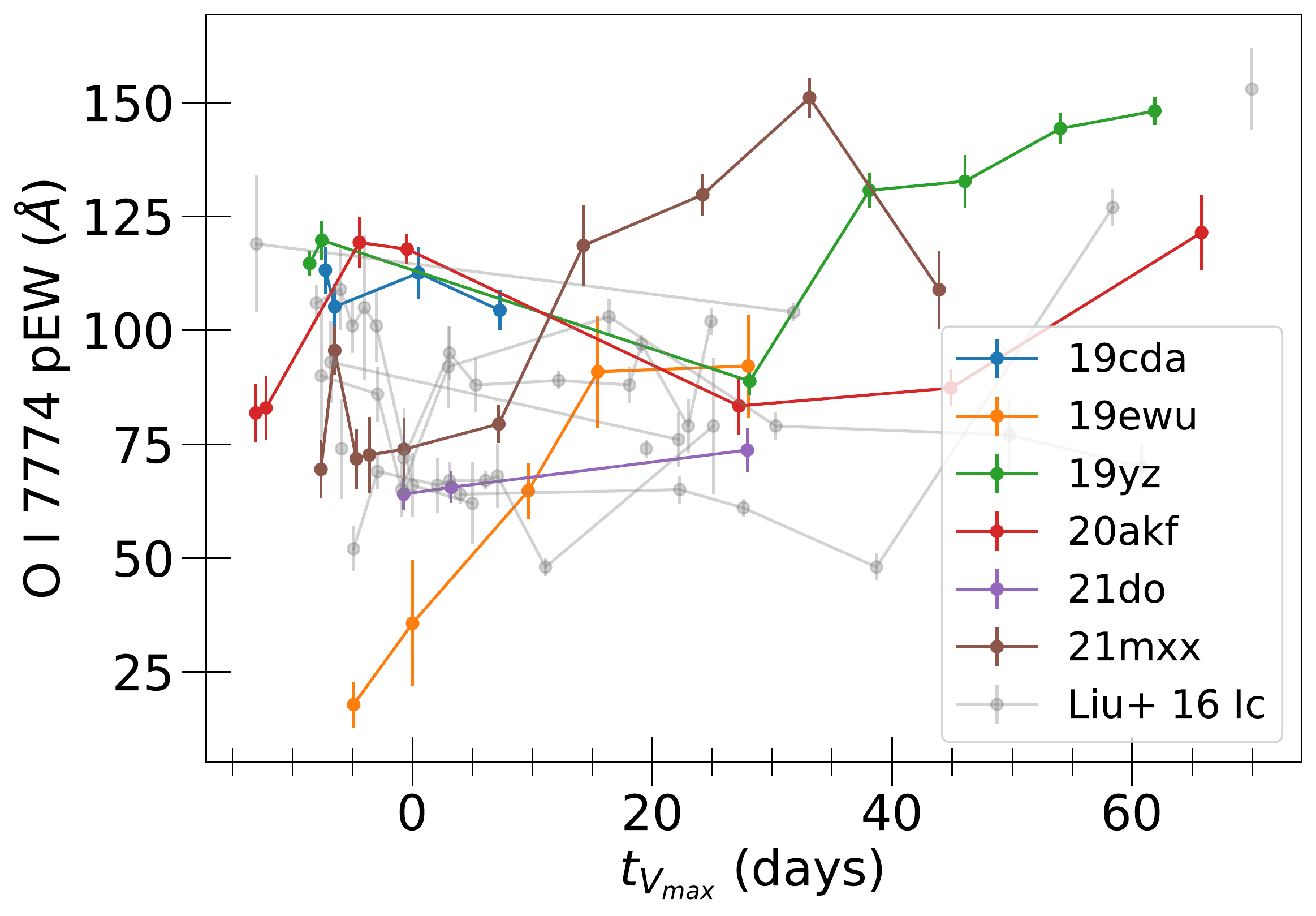}
 \includegraphics[scale=0.35]{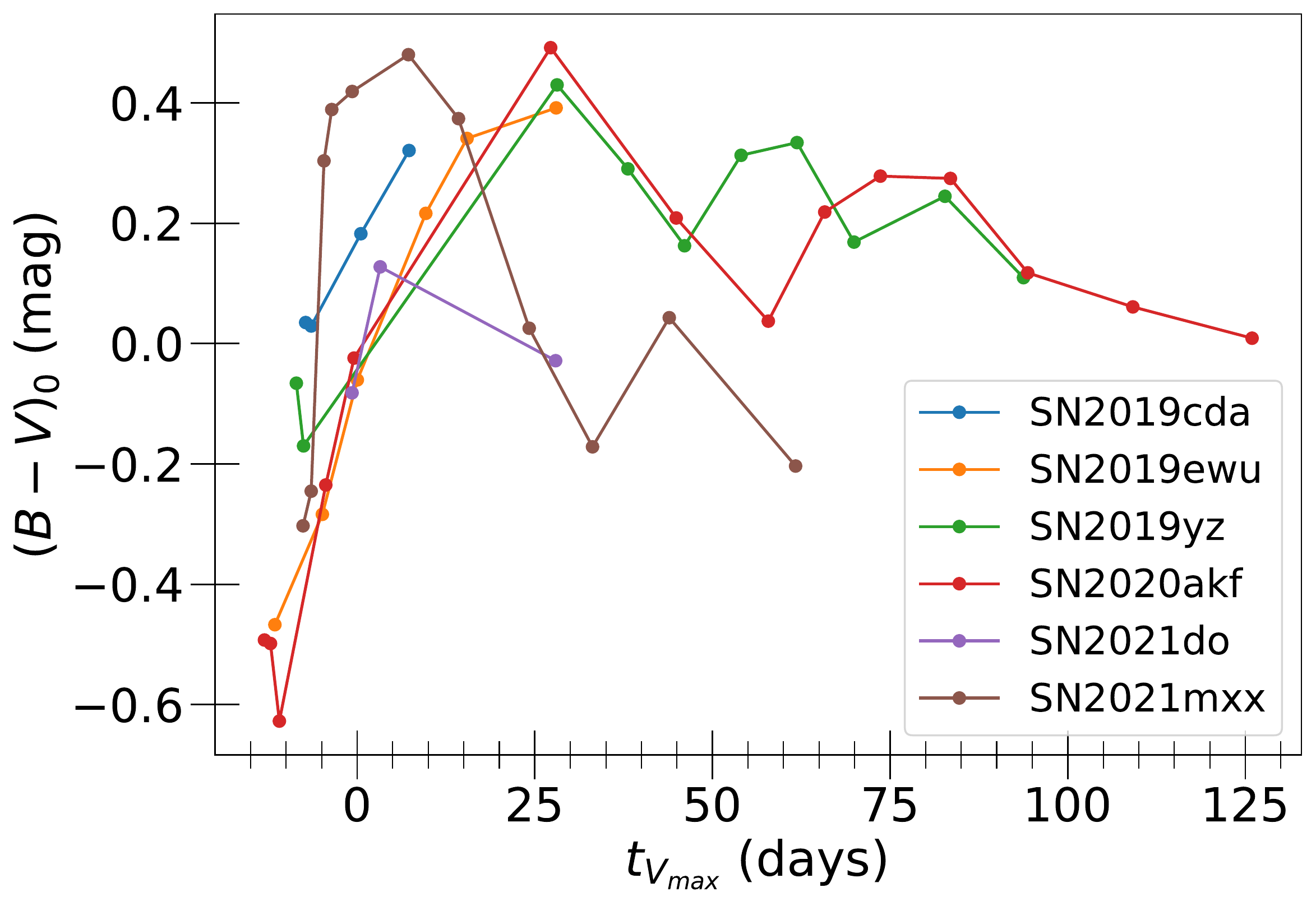}
    \caption{\textit{(Left)}: The O I 7774 absorption line pseudo equivalent width (pEW) as a function of time for the GSP SN Ic sample (colors) compared to the distribution of SNe Ic from the MGS (grey). SN~2019ewu has particularly weak Oxygen absorption at early times. \textit{(Right)}: (B-V) colors calculated directly from the GSP spectra using the ST-mag system. All spectra are de-reddened to correct for MW extinction, and SN~2019yz and SN~2021do are de-reddened to correct for host galaxy extinction. SN~2019ewu and SN~2020akf show particularly blue early-time spectra but different oxygen line strengths.}
     \label{fig:col}

\end{figure*}

\subsection{Classifying Young SNe Ic Spectra}
\label{sec:snid_disc}
Robustly classifying SN spectra as early as possible is crucial for identifying targets appropriate for follow-up observations (i.e. NIR or UV spectroscopy) particularly if follow-up takes time to schedule (i.e. $\sim$ 1 week for HST). In addition, science cases may require follow-up at or before the SN reaches maximum light. However, classifying young SN spectra is particularly challenging due to the scarcity of similarly young previously observed data. Template matching classification codes (e.g. SNID; \citealt{snid}, SuperFit; \citealt{howell2005gemini}) require previously observed young spectral templates of the same type in order to make robust classifications, and machine learning based algorithms (e.g. SESNPCA; \citealt{Williamson19}) require young training spectra. The MGS contains 23 SNe Ic and 259 spectra with $t_{V_{max}}<60$ days, but only 15 spectra have $t_{V_{max}}<-7$ days. The GSP SN Ic Sample presented in this paper includes 8 new spectra with $t_{V_{max}}<-7$ days and increases the total number of pre-maximum light spectra by 25\%. SNID templates for the SNe Ic presented in this paper are publicly available\footnote{\url{https://github.com/nyusngroup/SESNtemple}}.

Three of the six SNe Ic presented in this paper were initially misclassified using their earliest spectra. SN~2019ewu was classified as SN II due to the potential H$\alpha$ P-cygni profile and blue continuum. SN~2020akf and SN~2021mxx were initially misclassified as SNe Ia due to relatively strong potential Si II absorption along with the characteristic lack of H and He features. Using the SESN SNID templates from the current MGS, we find good spectral matches for these young misclassified spectra to a variety of stripped SNe types. SN~2021mxx matches multiple typical SNe Ic including SN~2007cl and SN~2004dn. We find good spectral matches for SN~2020akf to two stripped superluminous SNe (SLSNe Ic) iPTF13ajg \citep[][]{vreeswijk2014hydrogen} and SSS120810-23 \citep[][]{nicholl_slsne}. As shown by \citealt{liu2017_slsne}, SLSNe Ic and SNe Ic-bl exhibit similar velocities, and these matches reflect the high line velocities exhibited in SN~2020akf spectra that are discussed further in Section \ref{sec:velocity}. In addition, SN~2020akf exhibits a `W' shaped feature at early times near $\lambda4300\AA$ that is attributed to O II in SLSNe Ic and absent in SNe Ic \citep[][]{quimby2011hydrogen,gal2012luminous,nicholl2015diversity,mazzali2016spectrum,liu2017_slsne}.

SN~2019ewu is particularly interesting because it is the only SN Ic in the sample presented here that is misclassified as an H-rich SN II at early times. Figure \ref{fig:19ewu_types} shows a subset of the spectral time series for SN~2019ewu including spectral matches for each phase (colors). At $t_{V_{max}}=-12$ days, SNID identifies the typical SN IIP 1999em as a good spectral match to SN~2019ewu due to the P-cygni shaped feature near 6200$\AA$. As SN~2019ewu evolved, it more clearly became a stripped SN, though its high velocities yield best matches to SNe Ic-bl as it approached peak before matching to more typical SNe Ic at $t_{V_{max}}\geq 0$.

\begin{figure*}[t!]
 \centering
 \includegraphics[scale=0.5]{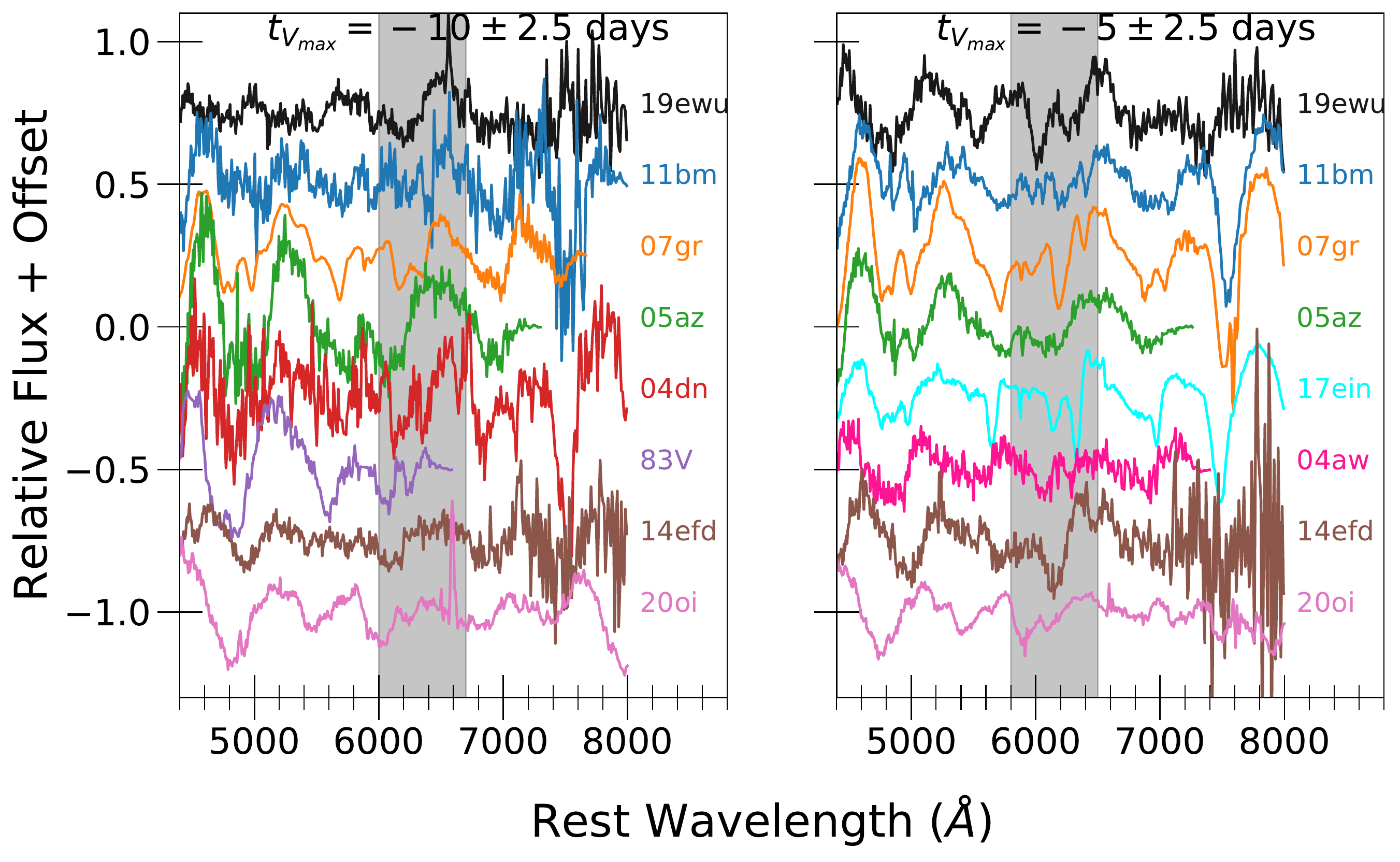}
 
    \caption{A comparison of the first two spectra of SN~2019ewu (black) to SNe Ic from the MGS at similar phase relative to V-band maximum (colors), as well as spectra from SN Ic 2020oi \citep[][]{rho_2020oi}. The spectra shown here are continuum removed by SNID to facilitate line shape comparison. The left figure highlights the broad P-Cygni feature in SN~2019ewu that has not been seen before in early time SN Ic spectra, while the right figure shows how the feature evolves one week later to develop a doublet shape.}
     \label{fig:19ewu_ic_comparison}

\end{figure*}

\subsection{High Velocity SNe Ic and the connection to SNe Ic-bl}
\label{sec:velocity}

SNe Ic-bl spectra have two distinguishing features compared to normal SNe Ic: extremely broad lines and systematically higher absorption velocities \citep[][]{modjaz_icbl}. However, previous sample studies of SNe Ic and Ic-bl spectra have identified a handful of outliers that appear to bridge these two classes. PTF12gzk is a SN Ic with typical narrow features but with high absorption velocities more consistent with SNe Ic-bl associated with gamma ray bursts \citep[][]{ptf12gzk,modjaz_icbl}. PTF2009dh was classified as SN Ic-bl due to early broad lines \citep[][]{prentice2016bolometric}, but was reclassified as a normal SN Ic \citep[][]{barbarino2021type}. Sample studies of SN Ic spectra measuring the absorption velocity of the Fe II 5169 line have shown that on average pre-maximum ejecta velocities are approximately 12,000 km/s and rarely exceed 15,000 km/s \citep[][]{modjaz_icbl,barbarino2021type}. However \citealt{modjaz_icbl} identify SN Ic 2004aw with extremely high pre-maximum velocities closer to 20,000 km/s, consistent with some SNe Ic-bl without associated GRBs \citep[][]{taubenberger2006sn}. Likewise for PTF10bip, \cite{modjaz20_icbl_host} concluded it was unclear whether the SN was Ic or Ic-bl due to the high velocities and blending they saw in the PTF spectra, and \cite{barbarino2021type} also concluded anomalously high Fe II 5169 velocity from the same PTF spectra which are not publicly available for comparison. Figure \ref{fig:fe_vel} shows the Fe II 5169 absorption velocity for the GSP SN Ic sample presented in this paper (colors) compared to the average velocity of the MGS (grey band). We find that the velocities of our sample are mostly consistent with those of the MGS with the exception of SN~2019ewu, which exhibits abnormally high velocities, particularly at early times ($t_{V_{max}}<0$) where $v_{abs}=18000$ km/s. In fact, early spectra of SN~2019ewu match with SN Ic-bl 2020scb (see Figure \ref{fig:19ewu_types}) and SN~2004aw. High velocity SNe Ic like SN~2019ewu may provide clues to the relationship between normal SNe Ic and SNe Ic-bl.

\subsection{Continuum Shape and Feature Strength}
\label{sec:color}

Figure \ref{fig:waterfalls} highlights the spectral diversity of the GSP SN Ic sample presented in this work. In particular, two aspects of the pre-maximum $(t_{V_{max}}<0)$ spectra are noteworthy: feature strength and continuum shape. The left panel of Figure \ref{fig:col} shows the O I 7774 pEW as a function of time for the GSP SN Ic sample (colors) compared to the MGS (grey). \citealt{modjaz_icbl} and \citealt{fremling_ptf_sesne} both showed that SNe Ic spectra have O I 7774 pEW in the range of 60-80$\AA$ at all phases, including before maximum light. However, we find that SN~2019ewu (already an outlier for its anomalously high velocities discussed in Section \ref{sec:velocity}) is also a strong outlier in this regard. We find the earliest spectrum ($t_{V_{max}}=-12$ days) does not have an identifiable O I 7774 feature, and when this feature does appear a week later, the pEW is less than 25$\AA$. This gives SN~2019ewu the weakest O I 7774 feature of all the SNe Ic analyzed in the MGS. The PTF SN Ic sample does have two SNe Ic with similarly weak O I 7774 early features \citep[see top panel of Figure 3; ][]{fremling_ptf_sesne}, but these SNe are not identified in the paper and the data behind the figure is not publicly available. The abnormally weak early O I 7774 feature in SN~2019ewu makes it an exception to the general trend first noted by \cite{matheson2001optical} that SNe Ic have stronger oxygen features than SNe Ib. By $t_{V_{max}}\approx20$ days the O I 7774 feature in SN~2019ewu has evolved to be consistent with the SN Ic class, highlighting the importance of early spectra for discovering new behavior in SNe Ic.

The right panel of Figure \ref{fig:col} shows the B-V colors using the ST-mag system calculated directly from the GSP SN Ic spectra. For colors calculated directly from spectra for SESNe with $t_{V_{max}}<20$ days there is a typical uncertainty of $\sim0.15$ mag \citep[][]{modjaz_14cfa} which cannot explain the range of colors observed in our sample. This uncertainty increases to $\sim0.37$ mag after $t_{V_{max}}>20$ days which may explain some of the oscillations that appear to be present in the color curves. SN~2019ewu distinguishes itself again from the majority of the SNe Ic in this sample with a particularly blue early continuum. We note that SN~2019ewu (orange) shows extremely similar continuum shape to SN~2020akf (red), but while SN~2019ewu exhibits relatively featureless early spectra, SN~2020akf shows typical early O I 7774 absorption with pEW$\approx80\AA$. Therefore a simple photospheric temperature difference is unlikely to explain the difference between the O I 7774 pEW strengths for these two SNe. More detailed radiative transfer modeling is required to investigate the differences between SN~2019ewu and SN~2020akf. Early blue relatively featureless continuua have also been observed in some standard SNe Ic-bl (SN~2014ad; \citealt{kwok_14ad}) as well as more exotic SNe Ic-bl (SN~2018gep; \citealt{pritchard_18gep}, iPTF16asu; \citealt{taddia2019analysis}), both of which are connected to the new class of Fast Blue Optical Transients (FBOTs; \citealt{inserra2019observational}). We note that the blue continuum exhibited by SN~2019ewu and SN~2020akf is also seen in the recently observed SN Ic 2022oqm \citep[][]{irani2022sn_oqm}.

\begin{figure*}[t!]
 \centering
 \includegraphics[scale=0.37]{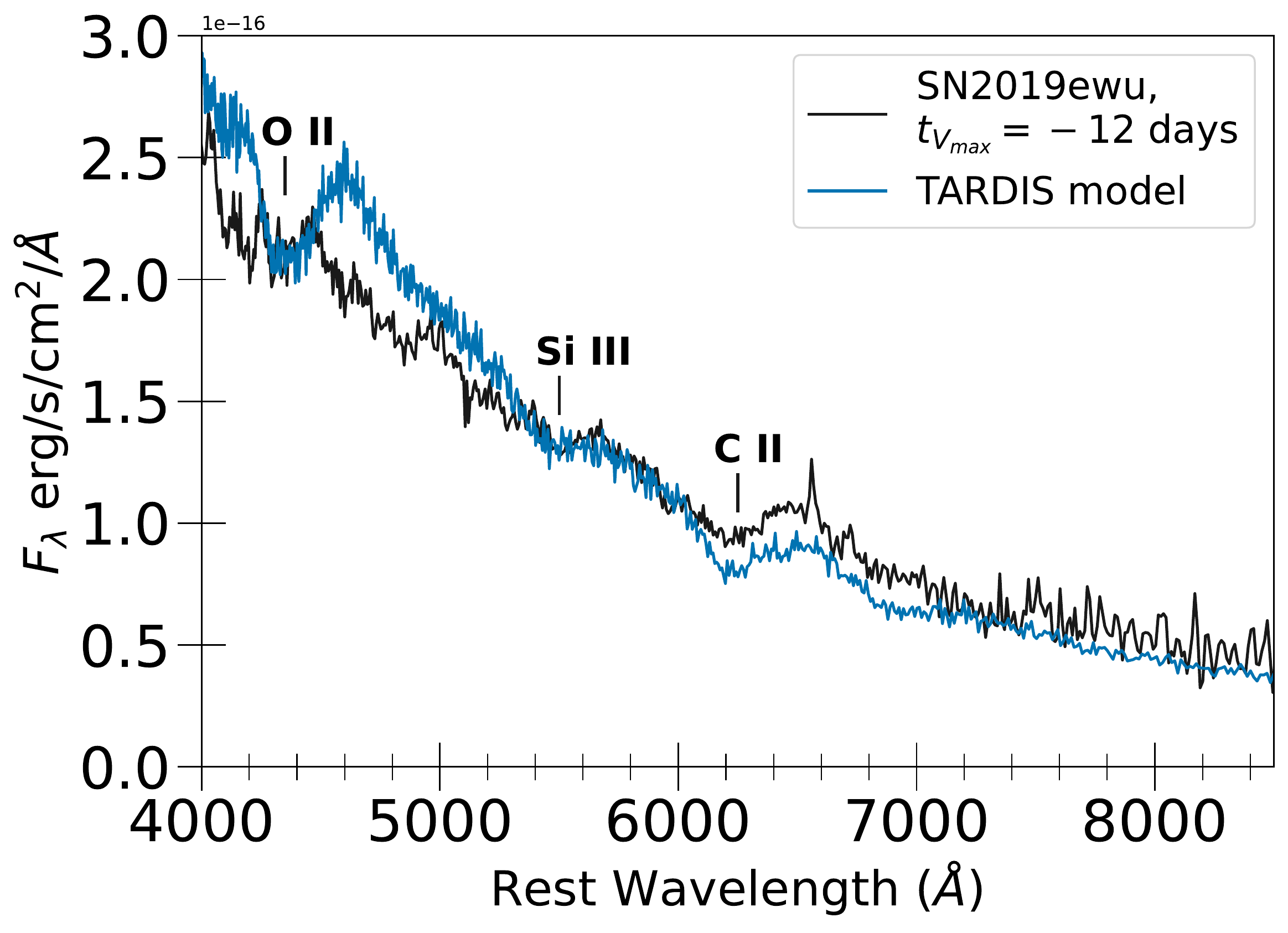}
 \includegraphics[scale=0.37]{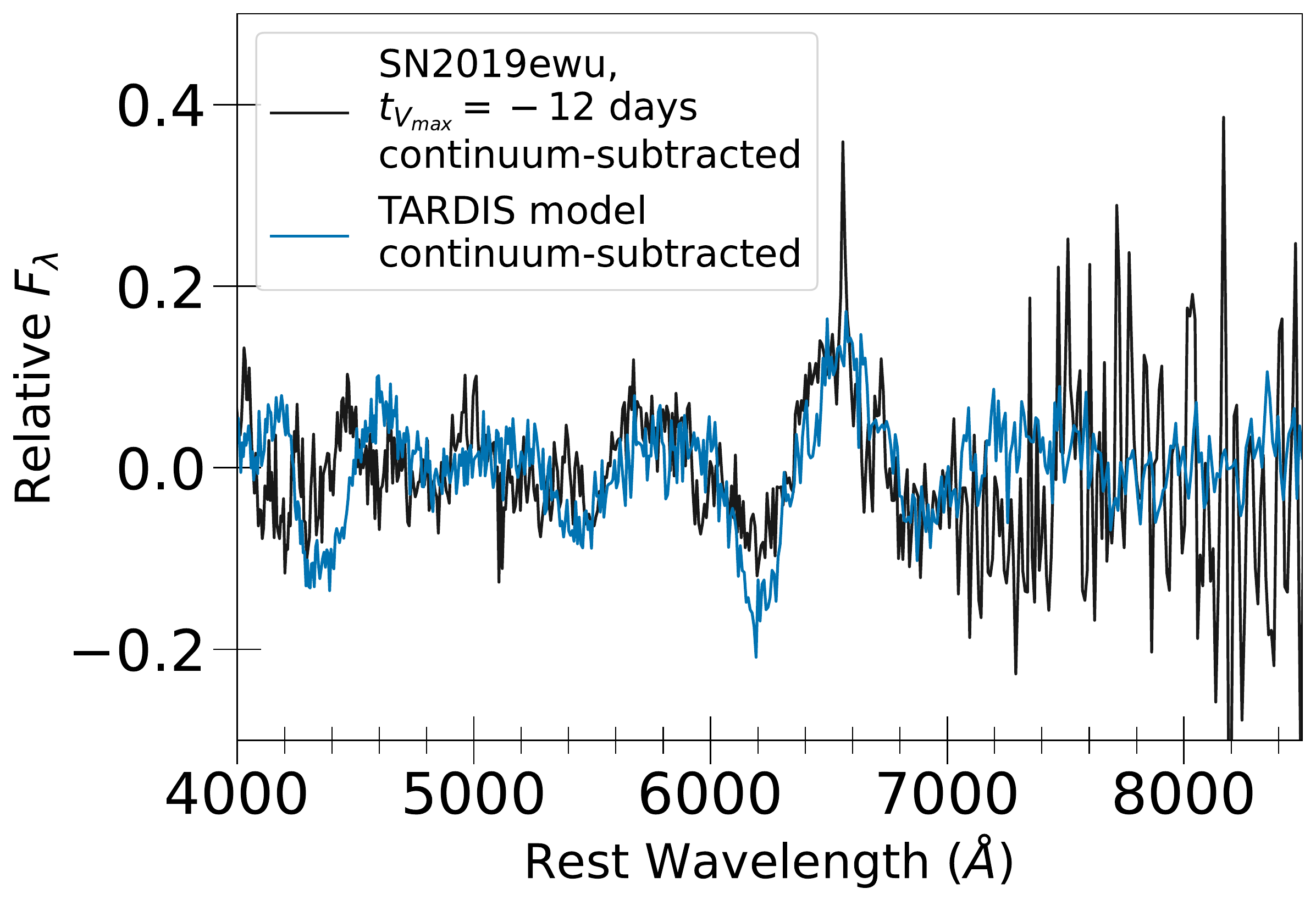}
    \caption{\textit{(Left)}: A TARDIS model (blue) for the first spectrum of SN~2019ewu (black). The primary features of the model are labeled (O II, Si III, C II). \textit{(Right)}: A comparison of the model and observed spectrum from the left panel with continuum removed by SNID to highlight the ability of the model to reproduce the lines seen in the data. The P-Cygni shaped feature in the SN~2019ewu spectrum can be produced in the model without any hydrogen.}
     \label{fig:19ewu_models}

\end{figure*}

\subsection{Comparing SN~2019ewu to Other SNe Ic}
\label{sec:19ewu}

SN~2019ewu is noteworthy as a SN Ic for its high velocities (see Section \ref{sec:velocity}) and early blue continuum with weak oxygen features (see Section \ref{sec:color}). In addition, one of its most puzzling characteristics is a similarity to SN II spectra at early times (see Section \ref{sec:snid_disc}). Figure \ref{fig:19ewu_ic_comparison} shows a comparison of the earliest two SN~2019ewu spectra to SNe Ic spectra at similar phases from the MGS as well as the recently discovered SN Ic 2020oi \citep[][]{rho_2020oi,gagliano2022early} which had particularly early spectral follow-up. In the left panel, the shaded region highlights that SN~2019ewu is the only SN Ic to show such a strong, clearly sinusoidal feature in the shaded region. This P-Cygni shaped feature coincides with the location for H$\alpha$ absorption and emission at 6563$\AA$, which explains the SNID template matches to SNe II spectra. However, SN~2019ewu does not exhibit any feature that could correspond to H$\beta$. A similar absorption feature seen in SN~2007gr is attributed to C II \citep[][]{valenti2008carbon}, but SN~2007gr does not exhibit the emission or P-cygni shape of the feature observed in SN~2019ewu. We note some similarity between SN~2005az, which was originally classified as both SN Ib and SN Ic \citep[][]{aldering05azatel,05azquimby}, but the emission part of the feature is too broad compared to SN~2019ewu. NIR spectra could help identify carbon lines (e.g. SN~2020oi; \citealt{rho_2020oi}, 2021krf; \citealt{21krf}), but no NIR spectra were obtained for SN~2019ewu.

The right panel of Figure \ref{fig:19ewu_ic_comparison} shows a spectral comparison for the subsequent SN~2019ewu spectrum a week later, where the broad P-Cygni feature has developed a doublet shape. We note that the evolution of SN~2019ewu in this respect is dramatically different from SN~2005az. The doublet feature shown in SN~2019ewu is seen in SN~2004aw to a weaker degree and possibly in SN~2007gr with a slower velocity structure. SN~2020oi also exhibits this doublet feature, but with slightly broader and weaker lines. Both SN~2004aw and SN~2007gr attribute this doublet feature to Si II $\lambda$6355 and C II $\lambda$6580 \citep[][]{taubenberger2006sn,valenti2008carbon}. SN~2007gr appears to be the best analog to SN~2019ewu for the feature highlighted in the shaded region of Figure \ref{fig:19ewu_ic_comparison}. We note that Figure \ref{fig:19ewu_ic_comparison} highlights the difference in O I 7774 feature strength. SN~2019ewu shows no O I 7774 in its earliest spectrum and a much weaker feature a week later in stark contrast to the pronounced O I 7774 feature in other carbon-rich SNe Ic SN~2007gr and SN~2017ein. SN~2022oqm also appears to show a similarly shaped C II absorption/emission feature early in its evolution that also evolves into a doublet attributed to carbon and silicon \citep[see Fig. 2][]{irani2022sn_oqm}. However there is a clear, broad, O I 7774 absorption feature in the SN~2022oqm SEDM spectrum at +6.46 day after explosion (which corresponds to $\sim$ 1 week before V-band maximum). SN~2019ewu is unique in having a combination of early carbon and a lack of O I 7774 absorption until $t_{V_{max}}=0$.

\subsection{TARDIS Models for SN~2019ewu}
\label{sec:19ewu_model}

Radiative transfer modeling is required in order to better understand the earliest spectrum ($t_{V_{max}}=-12$ days) of SN~2019ewu. It is particularly important to explain the P-Cygni shaped feature highlighted in the shaded region of the left panel of Figure \ref{fig:19ewu_ic_comparison} because this could be evidence of hydrogen. There has been some discussion in the literature of hydrogen potentially contributing to the spectral features of SN Ic 1994I \citep[][]{hachinger94Imodel,parrent2016line}, but there has not been a confident identification of hydrogen features in an SN Ic backed by modeling. The prevailing theory is that SNe Ic progenitors are highly stripped due to binary interactions \citep[][]{podsiadlowski92,yoon_ibc_prog_rev} or strong winds from massive single stars \citep[][]{crowther07,smartt09_rev} such that they have lost their hydrogen envelopes \citep[][]{schneider2020,woosley2020ibc}. 

Figure \ref{fig:19ewu_models} shows the results of modeling the earliest spectrum of SN~2019ewu with the public version of \gls{tardis}. The model assumes uniform abundances from elements typical of SNe Ic \citep[][]{dessart15,williamson_94i} including (He, C, N, O, Ne, Na, Mg, Si, S, Ca, and Fe-group elements). In particular, the model shown in Figure \ref{fig:19ewu_models} has a carbon abundance fraction of $\chi_{C}=0.35$ and does not include any hydrogen. The \gls{tardis} model file is publicly available\footnote{\url{https://github.com/tardis-sn/tardis-setups/tree/master/2022/williamson_19ewu}}. The left panel of Figure \ref{fig:19ewu_models} shows that the \gls{tardis} model (in blue) is overall a very good match to the data, although very slightly too blue, a difference that could be explained by a moderate amount of host extinction for SN~2019ewu. The right panel of Figure \ref{fig:19ewu_models} shows a comparison between SN~2019ewu and the model flattened by SNID to better compare the lines. The model reproduces the observed features in SN~2019ewu very well, including the P-Cygni shaped feature. This model does not include any hydrogen, and tracking the photon packet interactions from the simulation shows that more than 90\% of the absorption in the $\lambda6200$ feature is due to C II. Combined with the lack of observed H$\beta$ feature in the SN~2019ewu spectrum, we conclude that the P-Cygni shaped feature at $\lambda6200$ is most likely due to C II. In order to check whether hydrogen could contribute significantly to this feature, we have run a grid of simulations with the non-public \gls{tardis} version from \citealt{vogl_dil} which allows a more accurate treatment of hydrogen-rich SNe. We modified the baseline \gls{tardis} model from Figure \ref{fig:19ewu_models} to include a variety of hydrogen abundances up to 30\%. The results of these simulations (run including continuum processes and an NLTE treatment for H) show that the hydrogen contribution to the $\sim6200\AA$ feature is negligible compared to C II for the adopted composition and range of photospheric temperatures investigated. This is true even when H fractional abundance is increased to 30\%. In summary, we can be confident that the P-cygni profile seen in the -12 day spectrum of SN~2019ewu, even though it appears at wavelengths where SN II show H$\alpha$ and even though SNID matches the SN~2019ewu spectrum with those of Hydrogen rich SNe, is not due to H$\alpha$, but can be explained by C II.

\section{Summary}
\label{sec:summary}

In this paper we present a dataset of 49 spectra from 6 SNe Ic including a total of 17 spectra observed before V-band maximum. This dataset is publicly available and increases the number of available pre-maximum SN Ic spectra in the MGS by 25\%. These young spectra are critical to include as either templates or training data for supernova spectral classifiers so that robust classifications can be made closer to the date of explosion. This in turn will facilitate faster follow-up observations (e.g. HST UV spectroscopy) with less chance that an object has been misclassified (e.g. SN~2021yja; \citealt{sergiy_21yja}). In addition, extremely young spectra like the one observed for SN~2019ewu ($t_{V_{max}}=-12$ days) can provide important information on the outermost layers of the ejecta for SNe Ic, yielding constraints on stripping mechanisms and progenitor models. In particular we note that oxygen abundance in the ejecta has been tied to pre-explosion progenitor core mass \citep[][]{Dessart12} and recent work indicates that carbon and oxygen abundance gradients in the ejecta differ depending on whether the progenitor system is a massive single star versus a binary system \citep[][]{laplace_different_core,farmer_carbon}. We conduct a thorough analysis of the spectra presented here, including calculating Fe II 5169 velocities, O I 7774 feature strengths, and B-V colors. We also compare this new sample of SNe Ic to publicly available datasets like the MGS\citep[][]{modjaz2014optical, liu16, Williamson19}, PTF/iPTF sample \citep[][]{fremling_ptf_sesne}, and the CSP sample \citep[][]{shahbandeh_CSP_sesn}. We find that by all metrics, SN~2019ewu is an outlier SN Ic. SN~2019ewu exhibits extremely high ejecta velocities consistent with some SN Ic-bl spectra, while also displaying extremely weak O I 7774 features. SN~2019ewu is therefore a counterexample to the general trend that SNe Ic have stronger O I 7774 features than SNe Ib \citep[][]{matheson2001optical}. We show that SN~2019ewu exhibits peculiar early time behavior including a very blue continuum and a P-Cygni shaped feature near $\lambda6200\AA$ that evolves into a strong doublet a week later. Our extensive modeling efforts identify C II as the primary source of the early P-Cygni feature and not H$\alpha$. We therefore add SN~2019ewu to a growing subset of SNe Ic that exhibit particularly strong early carbon features. These carbon rich SNe Ic may be connected to the higher carbon yields derived from binary progenitor systems studied by \citealt{farmer_carbon} and warrant further study. Specifically, modeling of the entire SN~2019ewu spectral time series is called for to investigate the development of the strong doublet feature discussed in Section \ref{sec:19ewu} and to understand the abnormally weak early oxygen features.

\section{Acknowledgements}
This work makes use of data from the Las Cumbres Observatory network of telescopes. The LCO group is supported by NSF grants AST-1911225 and AST-1911151. This work was supported in part through the NYU IT High Performance Computing resources, services, and staff expertise. Marc Williamson is supported by the NASA Future Investigators in NASA Earth and Space Science and Technology grant (80NSSC21K1849). Christian Vogl was supported for part of this work by the Excellence Cluster ORIGINS, which is funded by the Deutsche Forschungsgemeinschaft (DFG, German Research Foundation) under Germany's Excellence Strategy-EXC-2094-390783311.

M.M. acknowledges support in part from NASA under the Swift GI program 1619152 (NASA grant No. 80NSSC21K0280), the Tess GI program G03267 (NASA grant No. 80NSSC21K0240), the ADAP program grant No. 80NSSC22K0486, the HST GO program HST-GO-16178.007, and teaching relief from a 19 Washington Square North Award.

This research made use of \textsc{tardis}, a community-developed software package for spectral
synthesis in supernovae \citep{2014MNRAS.440..387K, kerzendorf_wolfgang_2022_6662839}. The
development of \textsc{tardis} received support from GitHub, the Google Summer of Code
initiative, and from ESA's Summer of Code in Space program. \textsc{tardis} is a fiscally
sponsored project of NumFOCUS. \textsc{tardis} makes extensive use of Astropy and Pyne. Based on observations obtained with the Samuel Oschin 48-inch Telescope at the Palomar Observatory as part of the Zwicky
Transient Facility project. ZTF is supported by the National Science Foundation under Grant No. AST-1440341 and Grant No. AST-2034437 and a
collaboration including Caltech, IPAC, the Weizmann Institute for Science, the Oskar Klein Center at Stockholm University, the
University of Maryland, the University of Washington, Deutsches Elektronen-Synchrotron and Humboldt University, Los Alamos
National Laboratories, the TANGO Consortium of Taiwan, the University of Wisconsin at Milwaukee, and Lawrence Berkeley
National Laboratories, Trinity College Dublin, and IN2P3, France. Operations are conducted by COO, IPAC, and UW.

L.G. acknowledges financial support from the Spanish Ministerio de Ciencia e Innovaci\'on (MCIN), the Agencia Estatal de Investigaci\'on (AEI) 10.13039/501100011033, and the European Social Fund (ESF) ``Investing in your future'' under the 2019 Ram\'on y Cajal program RYC2019-027683-I and the PID2020-115253GA-I00 HOSTFLOWS project, from Centro Superior de Investigaciones Cient\'ificas (CSIC) under the PIE project 20215AT016, and the program Unidad de Excelencia Mar\'ia de Maeztu CEX2020-001058-M.

%\bibliography{williamson.bib}

\end{document}